\documentclass[journal]{IEEEtran}
\usepackage{amsfonts}
\usepackage{algorithmic}
\usepackage{algorithm}
\usepackage{array}
\usepackage[caption=false,font=normalsize,labelfont=sf,textfont=sf]{subfig}
\usepackage{textcomp}
\usepackage{stfloats}
\usepackage{url}
\usepackage{verbatim}
\usepackage{graphicx}
\usepackage{cite}
\usepackage{amsmath}
\usepackage{amsthm}
\usepackage{cuted}
\usepackage{amssymb}

\theoremstyle{remark}
\newtheorem{remark}{Remark}
\begin{document}
\title{CKM-Based Predictive Beamforming for Energy-Efficient Mobile Tracking in ISCAP System}
\author{Maryam Asadi-Ahmadabadi, S. Mohammad Razavizadeh, and Vahid Jamali
\thanks{Part of this paper has been presented at the IEEE 36th International Symposium on Personal, Indoor and Mobile Radio Communications (PIMRC), Istanbul, Turkiye, 01-04 September, 2025 \cite{11274715}.

M. Asadi-Ahmadabadi is with the School of Electrical Engineering, the Iran University of Science and Technology (IUST), Tehran, Iran
(e-mail: maryam\_asadi@elec.iust.ac.ir).

S. Mohammad Razavizadeh is with the School of Electrical Engineering, the Iran University of Science and Technology (IUST), Tehran, Iran (e-mail: smrazavi@iust.ac.ir).

Vahid Jamali is with the Department of Electrical Engineering
Technical University of Darmstadt,
Darmstadt, Germany (e-mail: vahid.jamali@tu-darmstadt.de).}}
\maketitle

\begin{abstract}
This paper proposes a channel-knowledge-map (CKM)-based predictive beamforming framework for integrated sensing, communication, and powering (ISCAP) of a mobile energy-constrained node in non-line-of-sight (NLoS) multipath environments. The CKM provides location-dependent channel information for both predictive beamforming and EKF-based tracking, while resolved path parameters estimated from the sensing echoes are used as observations. Based on the predicted node location, the transmit covariance and power-splitting ratio are jointly optimized to maximize wireless power-transfer efficiency subject to communication, sensing, transmit-power, and tracking-quality constraints. A finite-battery model and an adaptive inter-frame energy-management policy are further introduced to balance decoding energy consumption and battery preservation. The resulting non-convex fractional problem is solved using Dinkelbach's method, alternating optimization, and semidefinite relaxation. Numerical results show that the proposed framework maintains reliable tracking, satisfies the prescribed position-uncertainty requirement, achieves near-oracle wireless power-transfer performance after the initial tracking transient, and preserves more battery energy while meeting the required communication quality of service.
\end{abstract}
\section{Introduction}
The rapid growth of Internet of Things (IoT) networks and applications, including industrial monitoring, intelligent transportation, and autonomous systems, has increased the demand for reliable connectivity, environmental awareness, and sustainable energy supply. In this context, combining integrated sensing and communication (ISAC) with wireless power transfer (WPT) has led to the emergence of integrated sensing, communication, and powering (ISCAP) systems, which enable information transmission, spatial information acquisition, and wireless energy transfer within a unified framework by sharing signals, spectrum, and radio infrastructure \cite{SigModel_TWC_2024}. This capability is particularly important for battery-limited mobile IoT nodes, which rely on harvested energy for sustained operation and beam alignment for reliable transmission. Accordingly, maintaining beam alignment and satisfactory information and energy transfer performance under user mobility remains a fundamental design challenge.

In this regard, several studies have investigated beamforming design and optimization in ISCAP systems \cite{11146535, 11062362, 11357959, 11274715, 10556683}, targeting objectives such as data-rate maximization, sensing accuracy improvement, harvested-energy enhancement, or tradeoffs among multiple performance metrics. However, most conventional approaches design the transmit beam based on current channel information or user position. Under mobility, changes in the user position between channel estimation and data transmission can cause beam misalignment, reducing beamforming gain, communication quality, and harvested power. This issue is particularly critical for WPT, where efficient energy transfer requires accurate RF power focusing toward the intended receiver.

To address mobility-induced challenges, predictive beamforming uses historical information to predict the user’s future mobility state, angle, or channel and accordingly design or select the transmit beam \cite{9171304, 9246715, 9791349}. In ISAC systems, sensing information can further support user tracking by estimating parameters such as position, velocity, angle, delay, and Doppler, enabling a closed loop among sensing, tracking, and beamforming. Kalman filter (KF)- and extended KF (EKF)-based methods have been used to fuse sequential observations and predict user mobility states \cite{9171304, CKMTracking}. In \cite{fan2025mobility}, predictive beamforming was studied for tracking an energy receiver assuming known communication channel state information (CSI) in a line-of-sight (LoS) environment. In non-LoS (NLoS) scenarios, however, the observed channel parameters depend jointly on the device state and the locations and electromagnetic properties of environmental scatterers. Without accurate knowledge of the propagation environment, deriving an explicit mapping between the device state and sensing measurements is generally difficult, complicating channel prediction and predictive beamforming.

Channel knowledge maps (CKMs) provide a data-driven means of modeling location-dependent propagation characteristics \cite{CKM_Tutorial_2024} and thus offer a promising solution to this limitation. By mapping spatial coordinates to channel parameters, CKMs can predict the path gains, angles, delays, and Doppler shifts required for predictive transmission design. However, their integration into a mobile closed-loop ISCAP system raises several challenges. First, CKM predictions are uncertain due to map construction, calibration, and spatial interpolation errors. Second, online sensing measurements are noisy, with accuracy depending on the sensing SNR induced by the selected beamforming strategy. Third, beamforming affects not only instantaneous communication performance and harvested power but also tracking accuracy and, consequently, the reliability of future channel predictions. Explicitly accounting for these interdependencies is therefore essential for stable predictive beamforming.

To the best of our knowledge, existing ISCAP studies model the receiver energy requirement mainly through a minimum harvested-power constraint, which is insufficient for battery-constrained IoT devices. In practice, information decoding consumes considerable energy that increases with the transmission rate \cite{BATTERY_TWC_2014,BATTERY_TWC_2023,BATTERY_IDModel_2011}, so the battery state inherently limits the supportable rate. Moreover, in energy-harvesting nodes, the energy harvested in each frame determines the battery level and hence future operability. Thus, maximizing instantaneous performance without considering the battery state may cause rapid energy depletion and poor long-term sustainability, motivating the joint design of beamforming and energy management under mobility-dependent tracking uncertainty.

Motivated by these considerations, we propose a CKM-assisted predictive beamforming and mobility tracking framework, termed CKM-PBMT, for a mobile battery-constrained IoT device in an NLoS ISCAP system. At each frame, the base station (BS) predicts the device state and queries the CKM using the predicted position and velocity. The CKM output is partitioned into a channel-prediction subset for constructing the effective communication and sensing channels and a sensing-measurement subset for EKF-based tracking. The EKF observation covariance incorporates both CKM mapping uncertainty and beamforming-dependent online estimation error. Based on the predicted channel, battery state, and tracking covariance, the transmit beamforming vector and receiver power-splitting ratio are jointly optimized to maximize instantaneous WPT efficiency subject to communication-rate, harvested-power, transmit-power, and tracking-accuracy constraints.

The main contributions of this paper are summarized as follows.

\begin{itemize}

\item \textbf{CKM-assisted closed-loop ISCAP modeling:} A frame-based framework is developed to jointly model NLoS multipath propagation, communication, wireless power transfer, reflected-signal-based sensing, device mobility, and finite-battery dynamics. The sensing SNR of each path pair is related to the estimation-error variances of the angle, delay, and Doppler parameters, explicitly characterizing the dependence of tracking accuracy on the transmit beam.

\item \textbf{Uncertainty-aware CKM-EKF tracking:} The CKM serves both as a channel-prediction model and a nonlinear mapping from the device state to the sensing parameters. The proposed EKF distinguishes offline CKM mapping errors from online sensing-parameter estimation errors and combines them into an effective observation-noise covariance matrix, enabling the tracking-accuracy constraint to be directly incorporated into the transmission design.

\item \textbf{Battery- and tracking-uncertainty-aware energy management:} In addition to modeling rate-dependent decoding energy consumption and available battery capacity, an adaptive policy determines the receiver energy-consumption budget in each frame based jointly on the normalized battery level and posterior tracking uncertainty. Thus, when the battery is depleted or the predicted beam alignment is unreliable, a more conservative energy budget is allocated.

\item \textbf{Online beamforming design for WPT efficiency:} A per-frame WPT efficiency maximization problem is formulated over the transmit beamforming vector and power-splitting ratio. An alternating optimization algorithm based on Dinkelbach's method is developed, where the beamforming subproblem is reformulated as a semidefinite program and the power-splitting subproblem admits a closed-form solution. Moreover, under the stated conditions for a uniform linear array and the directional dependence of the problem functions, the semidefinite relaxation admits an equivalent rank-one optimal solution. The feasibility conditions, convergence behavior, computational complexity, and causal implementation of the proposed framework are also analyzed.

\end{itemize}
This work extends our previous conference paper \cite{11274715}, where we jointly optimized the transmit/receive beamforming vectors and power-splitting ratios to maximize the sum harvested energy in a multi-node ISCAP system, subject to communication- and sensing-SINR, harvested-power, and transmit-power constraints over Rician channels. In contrast, this work considers a mobile, battery-constrained IoT node in a multipath NLoS environment and develops a CKM-assisted predictive beamforming and mobility-tracking framework. The beamforming problem is further reformulated to maximize per-frame WPT efficiency under rate, harvested-power, transmit-power, and tracking-accuracy constraints.

The remainder of this paper is organized as follows. Section II presents the system model and beamforming problem formulation. Section III develops the CKM-assisted EKF tracking framework. Section IV solves the proposed optimization problem and analyzes the resulting algorithm. Section V presents the numerical results, and Section VI concludes the paper.

\textit{Notation:}
Scalars, vectors, and matrices are denoted by regular, bold lowercase, and bold uppercase letters, respectively. The real and complex number sets are denoted by $\mathbb{R}$ and $\mathbb{C}$, with $\mathbb{C}^{x\times y}$ representing the space of $x\times y$ complex matrices. For a matrix $\mathbf{X}$, $\mathbf{X}^{T}$, $\mathbf{X}^{H}$, $\mathbf{X}^{-1}$, $\mathrm{tr}(\mathbf{X})$, and $\mathrm{rank}(\mathbf{X})$ denote its transpose, conjugate transpose, inverse, trace, and rank, respectively, and $\mathbf{X}\succeq\mathbf{0}$ indicates positive semidefiniteness. The operators $\mathrm{diag}(\cdot)$, $\mathrm{blkdiag}(\cdot)$, and $\otimes$ denote diagonal construction, block-diagonal construction, and the Kronecker product, respectively, while $\mathbf{I}_{x}$ denotes the $x\times x$ identity matrix. Expectation with respect to a random variable $X$ is denoted by $\mathbb{E}
_{X}[\cdot]$. Moreover, $|\mathbf{x}|_{2}$ and $|x|$ denote the Euclidean norm and scalar magnitude, respectively, and $[x]^+\triangleq\max{x,0}$. Finally, $\mathcal{CN}(\boldsymbol{\mu},\boldsymbol{\Sigma})$ and $\mathcal{N}(\boldsymbol{\mu},\boldsymbol{\Sigma})$ denote circularly symmetric complex Gaussian and real Gaussian distributions with mean $\boldsymbol{\mu}$ and covariance $\boldsymbol{\Sigma}$, respectively.

\begin{figure}[!t]
    \centering
    \includegraphics[width=1.01\linewidth]{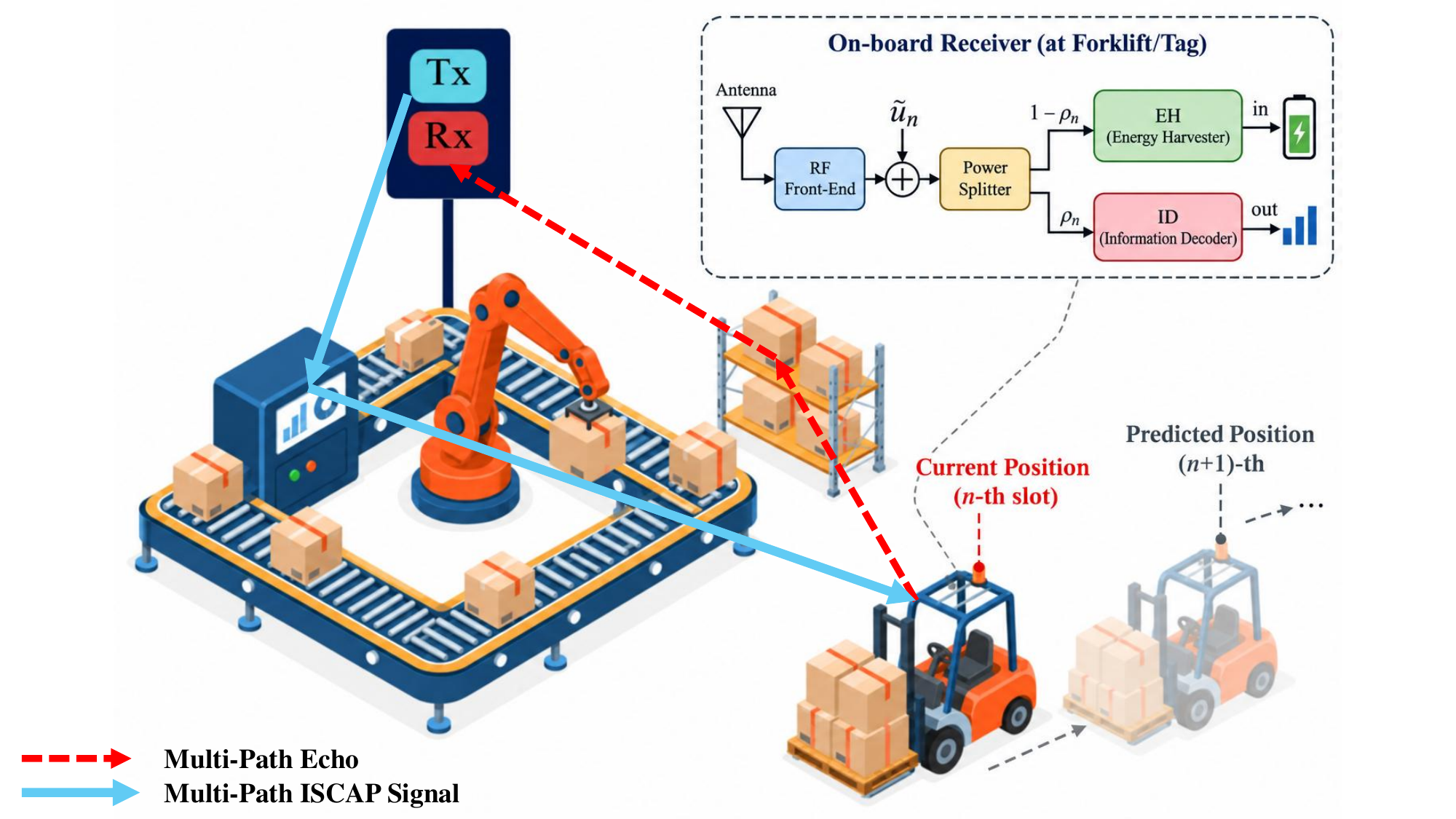}
    \caption{CKM-Enabled ISCAP for Mobile IoT Systems in NLOS Environments Under Battery Constraints.}
    \label{SystemModel}
\end{figure}
\section{System Model}
As shown in Fig. 1, we consider a multi-functional multiple-input single-output (MISO) downlink system comprising a BS and a mobile single-antenna IoT node. The BS employs a uniform linear array (ULA) with $N_t$ transmit and $N_r$ receive antennas and operates in full-duplex (FD) mode. The IoT node is equipped with a power splitter (PS) for simultaneous energy harvesting (EH) and information decoding (ID), along with a finite-capacity battery.

The BS transmits a multi-functional ISCAP signal to the IoT node while simultaneously receiving the corresponding echoes. These echoes are used to estimate the node location and velocity for tracking and beam selection toward efficient EH and ID. For sensing, the IoT node is modeled as a point target under a monostatic configuration.
\subsection{Channel Modeling}
In practical propagation environments, obstacles may cause both the BS-to-IoT node ISCAP signal and its echoes received at the BS to propagate through multiple paths. Accordingly, both the communication and sensing channels are modeled as multipath channels to capture realistic propagation conditions.

The DL communication channel, representing the forward propagation path, $\mathbf{h}_T(t,\tau) \in \mathbb{C}^{N_t \times 1}$, is modeled as
\begin{equation}
    \mathbf{h}_T(t, \tau) = \sum_{l=1}^{L} \alpha_{l} \mathbf a_{T}(\theta_l)\,
    \delta\!\left(\tau-\tau_{l}\right)
    e^{j2\pi\nu_{l} t},
    \label{DL channel}
\end{equation}
where $L$ denotes the total number of propagation paths in the forward link, and $\mathbf{a}_T(\cdot) \in \mathbb{C}^{N_t \times 1}$ denotes the transmit antenna steering vector. The parameters $\alpha_{l}$, $\tau_{l}$, $\theta_{l}$, and $\nu_{l}$ represent the complex path gain, propagation delay, AoD, and Doppler frequency shift associated with the $l$-th path, respectively.

Similarly, the reverse channel $\mathbf{h}_R(t,\tau) \in \mathbb{C}^{N_r \times 1}$, which is the reflected propagation path from the IoT node to the BS, can be expressed as
\begin{equation}
    \mathbf{h}_R(t,\tau) = \sum_{l'=1}^{L'} \alpha_{l'} \mathbf{a}_R(\theta_{l'})\,
    \delta\!\left(\tau-\tau_{l'}\right)
    e^{j2\pi\nu_{l'} t},
\end{equation}
where $L'$ denotes the total number of propagation paths in the reverse link, and $\mathbf{a}_R(\cdot) \in \mathbb{C}^{N_r \times 1}$ represents the receive antenna steering vector. The parameters $\alpha_{l'}$, $\theta_{l'}$, $\tau_{l'}$, and $\nu_{l'}$ denote the complex gain, AoA, propagation delay, and Doppler shift associated with the $l'$-th reverse-link path, respectively.

From the sensing perspective, the sensing response matrix (SRM) characterizes the overall round-trip propagation. Specifically, the BS transmits the signal to the IoT node through $\mathbf{h}_T(t,\tau)$, and the body of the IoT node reflects the received signal back to the BS through $\mathbf{h}_R(t,\tau)$  with a reflection coefficient $\varepsilon$. Accordingly, the SRM, $\mathbf{H}(t,\tau) \in \mathbb{C}^{N_r \times N_t}$, is given by

\begin{equation}
    \begin{aligned}
            \mathbf{H}(t,\tau)
    &=  \varepsilon \,  \mathbf{h}_R(t,\tau) \ast \mathbf{h}^{H}_T(t,\tau)  \\
    &= \varepsilon \sum_{l=1}^{L} \sum_{l'=1}^{L'}
    \alpha_{l,l'}
    \mathbf{a}_R(\theta_{l'})
    \mathbf{a}_T^H(\theta_{l})\,
    \delta\!\left(\tau-\tau_{l,l'}\right) \\
    & \times e^{j2\pi\nu_{l,l'} t}.
    \label{SRM}
    \end{aligned}
\end{equation}
Here, $*$ denotes convolution over the delay variable with matrix multiplication between the reverse-link response $\mathbf h_{\mathrm R}\in\mathbb C^{N_{\mathrm r}\times1}$ and the forward-link response
$\mathbf h_{\mathrm T}\in\mathbb C^{N_{\mathrm t}\times 1}$. 
$\varepsilon$ represents the radar cross section (RCS) of the IoT node.
Furthermore, $\alpha_{l,l'} = \alpha_l\alpha_{l'}$, $\tau_{l,l'} = \tau_l + \tau_{l'}$, and $\nu_{l,l'} = \nu_l + \nu_{l'}$ denote the cascaded complex path gain, round-trip delay, and round-trip Doppler shift, respectively.

Assuming half-wavelength spacing between adjacent antennas, the transmit and receive steering vectors corresponding to the $l$-th and $l'$-th paths are expressed as
\begin{equation}
\mathbf{a}_T(\theta_l)
=
\left[
1,\;
e^{j\pi \sin(\theta_l)},\;
\ldots,\;
e^{j\pi (N_t-1)\sin(\theta_l)}
\right]^T,
\end{equation}
and
\begin{equation}
\mathbf{a}_R(\theta_{l'})
=
\left[
1,\;
e^{j\pi \sin(\theta_{l'})},\;
\ldots,\;
e^{j\pi (N_r-1)\sin(\theta_{l'})}
\right]^T.
\end{equation}

\subsection{Signal Model}
We divide the time horizon into $N$ frames, indexed by $n \in \{1, 2, \dots, N\}$, each with a duration of $T_{\text{f}}$ seconds. We assume a quasi-static channel. Thus, the channels remain constant within each frame and vary between consecutive frames.

\subsubsection{Transmitted signal by the BS}
 Assuming linear beamforming, the transmitted ISCAP signal toward the IoT node at the $t$-th time instant within the $n$-th frame can be expressed as $\mathbf{x}_{n}(t) = \mathbf{w}_{n} s_{n}(t)$, where $\mathbf{x}_{n}(t) \in \mathbb{C}^{N_t \times 1}$ denotes the transmitted signal vector, $\mathbf{w}_{n} \in \mathbb{C}^{N_t \times 1}$ is the transmit beamforming vector, and $s_{n}(t)$ represents  the transmitted data symbol, which is normalized to have unit average power, i.e., $\mathbb{E}_s[|s_n(t)|^2]=1$.

The transmitted beamforming vector, constrained by the maximum power limit at the BS, is given by $P_{T,n} = \mathbb{E}_s[\left\| \mathbf{x}_{n}(t) \right\|^2] = \left\| \mathbf{w}_{n} \right\|^2 \leq P_{\text{max}}$,
where $P_{T,n}$ denotes the radiated power at the BS.
\subsubsection{Received signal by the IoT node}
Given the transmitted ISCAP signal, the received signal at the node over the DL communication channel described in (\ref{DL channel}) is
\begin{equation}
    \begin{aligned}
         y_{n}(t) 
               &= \sum_{l=1}^L \alpha_{l} \mathbf{a}_T^H(\theta_{l}) \mathbf{w}_{n} s_{n}(t-\tau_{l}) e^{j2\pi\nu_{l} t} + u_{n}(t),
    \end{aligned}
\end{equation}
where $u_{n} \sim \mathcal{CN}(0,\delta^2)$ denotes the additive white Gaussian noise (AWGN), with $\delta^2$ representing the noise power at the receiver antenna.

Similar to  \cite{11274715}, \cite{SigModel_TWC_2013}, \cite{BATTERY_TWC_2023}, we assume that the IoT node applies PS to the received signal $y_{n}(t)$ for simultaneous ID and EH. Let $\rho_{n} \in (0,1)$ denote the PS ratio allocated to the ID circuit of the node during frame $n$. Accordingly, the portion of the received signal forwarded to the ID circuit is given by $y_{n}^{\mathrm{ID}}(t) = \sqrt{\rho_{n}} \, y_{n}(t) + \tilde{u}_{n}(t)$,
where $\tilde{u}_{n}\sim \mathcal{CN}(0,\tilde{\delta}^2)$ denotes the additional noise introduced at the ID circuit of the node. Accordingly, the communication \text{SNR} (C-SNR) at the ID part of the node is given by (\ref{commSINR}).
\begin{figure*}[!t]
        \begin{equation}
    \begin{aligned}
        \Gamma_n^C(\mathbf{w}_n, \rho_n) &= \frac{\rho_{n} \, \mathbb{E}_s \{ |\sum_{l=1}^L \alpha_{l}\mathbf{a}_T^H(\theta_{l}) \mathbf{w}_{n} s_{n}(t-\tau_{l}) \, e^{j2\pi\nu_{l} t}|^2 \} }{\rho_n \, \mathbb{E}_u \{|u_{n}|^2 \} + \mathbb{E}_u \{|\tilde{u}_{n}|^2 \}}= \frac{\rho_{n} \, \Big( \sum_{l=1}^L 
        |\alpha_{l}|^2 \mathbf{a}_T^H(\theta_{l}) \mathbf{w}_{n} \mathbf{w}_{n}^H  \mathbf{a}_T(\theta_{l}) \Big)}{\rho_n \, \delta^2 +{\tilde{\delta}} ^2}.
        \label{commSINR}
    \end{aligned}
\end{equation}
\vspace{1mm}
\noindent\rule{\linewidth}{0.4pt}
\end{figure*}

To derive (\ref{commSINR}), we assume that the multipath components are resolvable in the delay domain, such that $B|\tau_l-\tau_k|\gg 1$, or equivalently, $R_s(\tau_l-\tau_k)\approx 0$ for $l\neq k$, where $B$ and $R_s(\cdot)$ denote the signal bandwidth and the waveform autocorrelation function, respectively. Hence, the delayed path components are approximately uncorrelated, and the cross terms vanish in the average received power. Meanwhile, the array is assumed to be spatially narrowband, i.e., $B\tau_{\mathrm{ap}}\ll 1$, where $\tau_{\mathrm{ap}}$ is the maximum differential propagation delay across the array aperture. Accordingly, the frequency dependence of the array response and the resulting beam-squint effect are neglected.

The remaining portion of the received signal is delivered to the EH circuit of the node during frame $n$, which is expressed as $y_{n}^{\mathrm{EH}}(t) = \sqrt{1 - \rho_{n}} \, y_{n}(t)$.

Consequently, the harvested energy at the EH circuit of the node during frame $n$ is given by
\begin{equation}
   \begin{aligned}
        E_n^{h}(\mathbf{w}_n, \rho_n) & = \zeta_{n} (1 - \rho_{n}) \, \mathbb{E}_s \!\left[ |y_{n}(t)|^2 \right] T_f\\
       & = \Big[ \sum_{l=1}^L  |\alpha_{l}|^2 \mathbf{a}_T^H(\theta_{l}) \mathbf{w}_{n} \mathbf{w}_{n}^H \mathbf{a}_T(\theta_{l}) + \delta^2 \Big]\\
       & \times \zeta_{n} (1 - \rho_{n}) \, T_f,
   \end{aligned}
   \label{EH}
\end{equation}
where $\zeta_{n} \in (0,1)$ denotes the energy conversion efficiency.
The wireless power transfer (WPT) efficiency for the node in frame $n$ is defined as the ratio of the harvested power $P_{n}^{h} = \frac{E_{n}^{h}}{T_f}$ at the node to the total power consumption at the BS, given by
\begin{equation}
    \eta_{n}=\frac{P_{n}^{h}(\mathbf{w}_n, \rho_n)}{P_{n}^{\text{tot}}(\mathbf{w}_n)},
\end{equation}
where $P_{n}^{\text{tot}} =  P_{T,n} /\varsigma+ P_{tx}^{\text{cir}}$, and $\varsigma \in(0,1]$ denotes the power amplifier efficiency. 
$P_{tx}^{\text{cir}}$ represents the fixed circuit power consumed by the transmitter front-end at the BS.

\subsubsection{Received echo signals by the BS}
When the BS transmits $\mathbf{x}_{n}(t)$ to the IoT node, the reflected echoes received at the BS through the SRM described in (\ref{SRM}) are given by
\begin{equation}
\begin{aligned}
        \mathbf{y}_{n}^{\text{BS}}(t)
     &= \varepsilon \sum_{l=1}^{L} \sum_{l'=1}^{L'}
    \alpha_{l,l'}
    \mathbf{a}_R(\theta_{l'})
    \mathbf{a}_T^H(\theta_{l})\,
    \mathbf{w}_{n} s_{n}(t-\tau_{l,l'}) e^{j2\pi\nu_{l,l'} t}\\
    &+\bar{\mathbf{u}}_n(t),
\end{aligned}
\end{equation}
where $\bar{\mathbf u}_n(t)\in\mathbb C^{N_{\mathrm r}\times1}$
denotes the effective AWGN vector at the sensing receiver, whose
total average power is given by
$\mathbb E\{\|\bar{\mathbf u}_n(t)\|^2\}=\bar{\delta}^{2}$. It also includes clutter interference and the residual self-interference resulting from full-duplex operation, which are considered as part of the background noise \cite{SigModel_TWC_2024}.

Under the same delay-resolvability and spatially narrowband conditions adopted for \eqref{commSINR}, the effective round-trip components associated with distinct path pairs are assumed to be approximately uncorrelated. Consequently, the corresponding cross terms vanish when computing the average received echo power, and the sensing SNR (S-SNR) is given by (\ref{SensingSNR}).

\begin{figure*}[!t]
    \begin{equation}
    \begin{aligned}
        \Gamma_n^S(\mathbf{w}_n) &= \frac{\mathbb{E}_s \{ || \varepsilon \sum_{l=1}^L \sum_{l'=1}^{L'} \alpha_{l,l'}  \mathbf{a}_R(\theta_{l'}) \mathbf{a}_T^H(\theta_{l})  \mathbf{w}_{n} s_{n}(t-\tau_{l,l'})e^{j2\pi\nu_{l,l'} t}||^2 \}}{\mathbb{E}_u \{ ||\bar{u}_{n}||^2\}  }\\
        & = \frac{ \sum_{l=1}^L \sum_{l'=1}^{L'} \varepsilon^2 \, |\alpha_{l,l'}|^2 \, ||\mathbf{a}_R^H(\theta_{l'})||^2 \, \mathbf{a}_T^H(\theta_{l}) \mathbf{w}_{n} \mathbf{w}_{n}^H \, \mathbf{a}_T(\theta_{l})}{\bar{\delta}^2}.
        \label{SensingSNR}
    \end{aligned}
\end{equation}
\vspace{1mm}
\noindent\rule{\linewidth}{0.4pt}
\end{figure*}
Furthermore, the S-SNR contribution associated with the
path pair $(l,l')$ is defined as
\begin{equation}
\Gamma^{S}_{(l,l'),n}(\mathbf w_n)
=
\frac{
|\varepsilon|^2|\alpha_{l,l'}|^2
\|\mathbf a_{R}(\theta_{l'})\|^2
\mathbf a_{T}^{H}(\theta_l)
\mathbf w_n\mathbf w_n^{H}
\mathbf a_{T}(\theta_l)
}{
\bar{\delta}^{2}
}.
\end{equation}
Accordingly, the total S-SNR in (\ref{SensingSNR}) can
be expressed as
$\Gamma_n^{S}(\mathbf w_n)
=
\sum_{l=1}^{L}\sum_{l'=1}^{L'}
\Gamma^{S}_{(l,l'),n}(\mathbf w_n)$.
In the considered NLoS scenario, the AoD, AoA, propagation delay, and Doppler shift associated with each resolvable path pair $(l,l^{\prime})$ are estimated from the echo received at the BS. Rather than retaining the high-dimensional raw echo waveform, we abstract the sensing front-end through these estimated physical parameters, which provide a compact representation of each resolvable component. Accordingly, the sensing-parameter measurement vector for path pair $(l,l^{\prime})$ is defined as $\mathbf{z}_{(l,l^{\prime}),n}
=
\left[
\theta_{l,n},
\theta_{l^{\prime},n},
\tau_{(l,l^{\prime}),n},
\nu_{(l,l^{\prime}),n}
\right]^{T}$.

These parameters can be extracted using conventional techniques such as matched filtering and MUSIC. Although the same forward-link angle $\theta_{l,n}$ or reverse-link angle $\theta_{l^{\prime},n}$ may appear in multiple path-pair blocks, each resolvable round-trip component provides a separate noisy estimate, and the estimation errors of distinct path-pair components are assumed independent. Hence, the aggregate sensing-parameter measurement vector is
\begin{equation}
\mathbf{z}_{n}
=
\left[
\mathbf{z}_{(1,1),n}^{T},
\dots,
\mathbf{z}_{(L,L^{\prime}),n}^{T}
\right]^{T}
\in
\mathbb{R}^{4LL^{\prime}\times 1}.
\label{Zn}
\end{equation}

Since these parameters are estimated from the received echoes, their accuracy depends on the received S-SNR. Under AWGN, the Fisher information scales with SNR, whereas the corresponding Cramér--Rao lower bound (CRLB) scales inversely with it \cite[Ch.~3]{kay1993fundamentals}. Thus, in the high-SNR regime and for efficient or approximately unbiased estimators, for
$\Omega_i
\in
\left\{
\theta_l,
\theta_{l^{\prime}},
\tau_{l,l^{\prime}},
\nu_{l,l^{\prime}}
\right\}$,
we adopt \cite{fan2025mobility}
\begin{equation}
\sigma_{\Omega_i}^{2}
=
\frac{c_{\Omega_i}}
{\Gamma_{(l,l^{\prime}),n}^{S}},
\label{sigomega}
\end{equation}
where $\Gamma_{(l,l^{\prime}),n}^{S}$ denotes the S-SNR of path pair $(l,l^{\prime})$, and $c_{\Omega_i}$ is a parameter-dependent scaling factor determined by the transmitted waveform, signal bandwidth, matched-filter gain, antenna-array geometry, and observation duration.

\subsection{Battery-Limited Modeling}
We consider the IoT node to be equipped with a battery of finite capacity, denoted by $E_{\text{max}}$, and an EH source that enables the node to dynamically collect energy from RF signals. The battery energy level decreases in accordance with the circuit power consumption and the decoding operations of the downlink received data \cite{BATTERY_TWC_2023}, \cite{BATTERY_TWC_2014}. The EH circuit allows the node to recharge its battery by collecting the required energy from received RF signal. We assume that the EH system adopts the harvest-store-use strategy \cite[Sec. III]{EHModel_2015},
wherein the harvested energy in a given frame is available only in subsequent frames. Thus, the battery dynamics of the node at the beginning of the frame $(n +1)$ is given by
\begin{equation}
    E_{n+1}=\text{min} \{E_{\text{max}},[E_n-E_n^u]^+ +E_n^h\},
    \label{Ctilt}
\end{equation}
where $E_n^h$ denotes the harvested energy defined in (\ref{EH}). To conservatively avoid battery overflow, the harvested energy is upper-bounded by the battery headroom available at the beginning of frame $n$, i.e.,
\begin{equation}
    E_n^{h_{\text{max}}} = E_{\text{max}} - E_n.
    \label{Emax}
\end{equation}
Furthermore, $E_n^u$ denotes the energy consumed by the node during frame $n$, which is modeled as
\begin{equation}
    E_n^u=T_{\text{f}}(P_{rx}^{\text{cir}}+P_n^{\text{dec}}),
    \label{useenergy}
\end{equation}
To regulate the receiver-side energy expenditure across frames, let $E_n^{u,\mathrm{bud}}$ denote the allowable energy-consumption budget assigned at the beginning of frame $n$. Accordingly, the actual energy consumed during frame $n$ must satisfy
\begin{equation}
0\leq E_n^u\leq E_n^{u,\mathrm{bud}}\leq E_n.
\label{energy_budget_constraint}
\end{equation}
Classical precoder designs for macro-cell systems typically assume sufficient receiver energy for reliable reception and decoding. This assumption may not hold for battery-limited IoT devices, where decoding energy is non-negligible and can be comparable to the RF reception power \cite{BATTERY_TWC_2014}. Hence, the transmission rate must account for the receiver's available energy, as the battery state inherently limits the achievable rate even under favorable channel conditions.

Although no established analytical model for decoding energy consumption exists \cite{BATTERY_TWC_2023,BATTERY_TWC_2014}, decoder power consumption, denoted by $P^{\text{dec}}(R)$, is generally assumed to increase with the data rate $R$. Higher rates require processing more bits per unit time and may increase the decoding iterations needed to meet a target bit error rate (BER), thereby increasing computational load and power consumption. For a fixed target BER, decoding energy can therefore be modeled as a function of the data rate \cite{BATTERY_TWC_2014}.
Various models have been proposed, including linear \cite{BATTERY_TWC_2023, BATTERY_Linmodel_2016} and exponential \cite{BATTERY_Expmodel_2010} models. Following \cite{BATTERY_TWC_2023,BATTERY_Linmodel_2016}, we adopt the linear model
\begin{equation}
P_n^{\text{dec}}=\mu R_n,
\label{declinear}
\end{equation}
where the decoder power consumption increases linearly with the data rate. $\mu$ models the decoder efficiency. By combining (\ref {useenergy}) and (\ref{declinear}) with the energy-budget constraint in \eqref{energy_budget_constraint}, the maximum allowable
rate is obtained as
\begin{equation}
    R_n \le R_n^{\text{max}} \triangleq \frac{1}{\mu}(\frac{E_n^{u,\mathrm{bud}}}{T_f}-P_{rx}^{\text{cir}}).
    \label{data-rate}
\end{equation}
\subsection{Allowed Energy Allocation Among Frames}

In battery-limited EH-enabled systems, resource allocation affects both instantaneous performance and the energy available for future transmissions. Hence, short-term optimization may rapidly deplete the battery and degrade long-term performance. Since per-frame energy consumption is coupled with EH and battery evolution, an energy-budget mechanism is required to balance instantaneous performance and long-term sustainability.

To this end, \cite{BATTERY_TWC_2014} proposes a fixed policy that allocates a constant fraction of the available battery energy to the next-frame consumption budget. Let $E_{n+1}^{u,\mathrm{bud}}$ denote this budget $E_{n+1}^{u,\mathrm{bud}}=\psi E_{n+1}$,
where $\psi\in(0,1)$ is a constant allocation coefficient. Although this policy captures the trade-off between instantaneous performance and battery lifetime, its fixed coefficient cannot adapt to time-varying battery and tracking conditions. More advanced approaches, such as \cite{BATTERY_TWC_2023}, dynamically regulate receiver-side energy consumption according to the system state, including the battery level, channel conditions, and queue status, but do not explicitly exploit the tracking uncertainty available in ISAC systems.

In contrast, the proposed framework uses the EKF tracking information to adapt the energy-consumption budget in each frame by replacing $\psi$ with
\begin{equation}
\psi_{n+1}
=\frac{1}
{1+\beta_1\mathrm{tr}(\bar{\mathbf{\mathbf{\Phi}}}_n)+\beta_2/\bar{E}_n},
\label{adaptive_budget_factor}
\end{equation}
where $\beta_1,\beta_2>0$ control the sensitivity to tracking uncertainty and battery level, respectively. The normalized posterior covariance is $\bar{\mathbf{\mathbf{\Phi}}}_n = \mathcal{T}^{-1}\mathbf{\mathbf{\Phi}}_n\mathcal{T}^{-T}$, 
where $\mathcal{T}=\operatorname{diag}(\epsilon_{p_x},\epsilon_{p_y},\epsilon_{v_x},\epsilon_{v_y})$, with $\epsilon_p$ and $\epsilon_v$ denoting reference position and velocity errors, and $\mathbf{\mathbf{\Phi}}_n$ the posterior tracking-error covariance (MSE). This normalization ensures consistent scaling of position and velocity components. Moreover, $\bar E_n=E_n/E_{\max}$ is the normalized battery level.

A larger $\mathrm{tr}(\bar{\mathbf{\mathbf{\Phi}}}_n)$ or a lower $E_n$ decreases $\psi_{n+1}$, yielding a more conservative energy budget under higher tracking uncertainty or limited battery reserves. Accordingly,
\begin{equation}
E_{n+1}^{u,\mathrm{bud}}=
\psi_{n+1}E_{n+1}.
\label{adaptive_energy_budget}
\end{equation}

\subsection{Problem Formulation}
This work develops a WPT-efficient resource allocation framework that satisfies node tracking accuracy and battery constraints. Specifically, the real-time WPT efficiency is maximized per frame by jointly optimizing the transmit beamforming vectors and receive PS ratios, subject to QoS constraints on data rate, EH, and sensing performance.

The instantaneous tracking performance is characterized by the posterior MSE, $\mathrm{tr}(\bar{\mathbf{\mathbf{\Phi}}}_n)$, which implicitly depends on the beamforming vector $\mathbf{w}_n$ through \eqref{SensingSNR} and \eqref{sigomega}. Accordingly, the per-frame optimization problem is formulated as follows

\begin{equation}
\begin{aligned}
\mathrm{P0}: \quad \max_{\mathbf{w}_n, \rho_n} \quad & \eta_{n}(\mathbf{w}_n, \rho_n)=\frac{\hat{P}_{n}^{h}(\mathbf{w}_n, \rho_n)}{P_{n}^{\text{tot}}(\mathbf{w}_n)} \\
\textrm{s.t.} \quad 
& \text{C}_1: \|\mathbf{w}_n\|^2 \leq P_{\text{max}}, \\
& \text{C}_2: R_n^{\text{min}} \leq \hat{R}(\mathbf{w}_n,  \rho_n) \leq R_n^{\text{max}}, \\
& \text{C}_3: P_{n}^{h_\text{min}} \leq \hat{P}_n^{h}(\mathbf{w}_n, \rho_n) \leq P_{n}^{h_\text{max}}, \\
& \text{C}_4: \mathrm{tr}(\hat{\bar{\mathbf{\mathbf{\Phi}}}}_n(\mathbf{w}_n) ) \leq \mathrm{tr}(\bar{\mathbf{\Phi}})_{\text{max}}, \\
& \text{C}_5: 0 < \rho_n < 1.
\end{aligned}
\label{p0_original}
\end{equation}

Here, $P_{\text{max}}$ is the maximum transmit power. The predicted transmit rate $\hat{R}_n$ is constrained by $C_2$ to satisfy the minimum QoS requirement $R_n^{\text{min}}$, while its upper limit follows from~\eqref{data-rate}. Similarly, $\hat{P}_n^{h}$ denotes the predicted harvested power, and $C_3$ imposes its minimum requirement and the upper bound $E_n^{h_{\text{max}}}/T_f$ from~\eqref{Emax}. Constraint $C_4$ limits the normalized predicted posterior MSE, $\mathrm{tr}(\hat{\bar{\mathbf{\Phi}}}_n)$, by the predefined threshold $\mathrm{tr}(\bar{\mathbf{\Phi}})_{\max}$. We assume $\hat{\Gamma}_n^{C}>0$ and $\hat{P}_n^{h}>0$, yielding $0<\rho_n<1$.

Solving $\mathrm{P0}$ predictively requires the upcoming-frame channel parameters and tracking-error covariance. The EKF predicts the node state, whose location is then used to query the offline CKM for path angles, gains, and delays. The predicted angles and gains determine the communication and sensing quantities in $\mathrm{P0}$, while the angles and delays support CKM-assisted state tracking.

\section{CKM-Assisted IoT Node Tracking}
To maintain beam alignment, the BS tracks the IoT node using channel parameters extracted from the received echoes. Since the state and measurement models are nonlinear, an EKF is employed for closed-loop tracking and multi-source data fusion \cite{Kalman1960}.
Assuming a BS located at $(0,0)$, the node has initial position $(x_0,y_0)$ and velocity $(\dot{x}_0,\dot{y}_0)$. Its state at frame $n$ is defined as
$\mathbf{s}_{n}=\big[x_n,y_n,\dot{x}_n,\dot{y}_n\big]^{\mathrm{T}}$, where $(x_n,y_n)$ and $(\dot{x}_n,\dot{y}_n)$ denote the node position and velocity, respectively. The corresponding state-transition and measurement models are developed next.

CKM characterizes the location-dependent relationship between node positions and channel knowledge through a site-specific channel-parameter map \cite{CKM_Tutorial_2024}. In the proposed NLoS framework, this map is constructed offline as a position-indexed database of multipath characteristics and is defined as
\begin{equation}
\mathcal{M}( \mathbf{p}_n) = \{\theta_{l,n},\theta_{l^{\prime},n}, |\alpha_{l,n}|, |\alpha_{l',n}|,\tau_{l,n}, \tau_{l',n},\boldsymbol{\kappa}_{l,n}, \boldsymbol{\kappa}_{l',n} \},
\end{equation}
where $\mathcal{M}(\cdot)$ denotes the learned CKM, and $\mathbf{p}_n=[x_n,y_n]^{T}$ represents the IoT node position at frame $n$. The vector
$\boldsymbol{\kappa}_{l,n}
=
[\kappa_{x,l,n},\kappa_{y,l,n}]^{T}$
denotes the location-dependent geometric Doppler coefficient of the $l$-th propagation path, which is determined by the path geometry and characterizes the projection of the IoT node velocity onto the corresponding propagation direction.

Let $L_l(\mathbf{p}_n)$ denote the physical length of the $l$-th path. The geometric Doppler coefficient is obtained from the spatial variation of the path length as
$\boldsymbol{\kappa}_{l,n} = \frac{1}{\lambda} \nabla_{\mathbf{p}_n}L_l(\mathbf{p}_n)$.
Accordingly, the instantaneous Doppler shift of the $l$-th path is reconstructed online as
$\nu_{l,n}
=
\boldsymbol{\kappa}_{l,n}^{T}\dot{\mathbf{p}}_n$,
where $\dot{\mathbf{p}}_n$ denotes the IoT node velocity. This formulation indicates that the CKM only stores the environment-dependent geometric information offline, while the instantaneous Doppler shift is obtained online using the current node velocity.

For a sensing path pair $(l,l^{\prime})$, the corresponding propagation delay and Doppler shift are expressed as
$\tau_{(l,l^{\prime}),n}
=
\tau_{l,n}+\tau_{l^{\prime},n}$
and
$\nu_{(l,l^{\prime}),n}
=
(\boldsymbol{\kappa}_{l,n}
+
\boldsymbol{\kappa}_{l^{\prime},n})^{T}
\dot{\mathbf{p}}_n$,
respectively.

The CKM output is used for both channel prediction and sensing-parameter prediction. For predictive beamforming and SNR evaluation, the required subset is
\begin{equation}
\mathcal{M}_{\mathrm{ch}}
\left(
\mathbf{p}_n
\right)
=
\left\{
\theta_{l,n},
\theta_{l^{\prime},n},
|\alpha_{l,n}|,
|\alpha_{l^{\prime},n}|
\right\},
\end{equation}
For sensing-assisted tracking, the CKM entries required to construct the predicted sensing parameters are collected in the tracking-related subset
\begin{equation}
\mathcal{M}_{\text{trck}}\left(\mathbf{p}_n\right)
= \{\theta_{l,n},\theta_{l^{\prime},n},\tau_{l,n}, \tau_{l',n},\boldsymbol{\kappa}_{l,n}, \boldsymbol{\kappa}_{l',n} \}.
\label{M_trck}
\end{equation}

\subsection{State Prediction Model}
The node motion follows a constant-velocity (CV) model within the time horizon of interest, assuming approximately constant velocity between consecutive frames, with acceleration modeled as random process noise \cite{EKF_RA_2024}
\begin{equation}
\mathbf{s}_{n} = \mathbf{F}\, \mathbf{s}_{n-1} + \mathbf{v}_{n-1},
\end{equation}
where $\mathbf{F}$ denotes the transition matrix given by $\mathbf{F} = [\mathbf{I}_2 , T_f \mathbf{I}_2; \mathbf{0}_2 , \mathbf{I}_2]$.
Here $\mathbf{I}_2$ is the $2 \times 2$ identity matrix. $\otimes$ is the Kronecker product.
The process noise vector $\mathbf{v}_{n} \sim \mathcal{N}(\mathbf{0}, \mathbf{V}_{n})$ represents the motion uncertainty of the node at the $n$-th frame, where the process noise covariance matrix is defined as $\mathbf{V}_{n} = \tilde{\sigma} [\frac{1}{3} T_f^3 \mathbf{I}_2 , \frac{1}{2} T_f^2 \mathbf{I}_2; \frac{1}{2} T_{\text{f}}^2 \mathbf{I}_2, T_{\text{f}} \mathbf{I}_2]$ \cite{EKF_RA_2015}.
Here $\tilde{\sigma}$ denotes the level of the process noise of the node.

\subsection{Sensing-Parameter Measurement Model}

As defined previously, the sensing-parameter vector is estimated online at the BS from the received echo signals. The corresponding measurement model is expressed as \cite{EKF_RA_2024}
\begin{equation}
\mathbf{z}_n
=
f(\mathbf{s}_n)
+
\tilde{\mathbf{v}}_n,
\end{equation}
where $f(\cdot)$ denotes the nonlinear state-to-measurement function, and $\tilde{\mathbf{v}}_n
=
[
\tilde{\mathbf{v}}_{(1,1),n}^{T},
\dots,
\tilde{\mathbf{v}}_{(L,L^{\prime}),n}^{T}
]^{T}$
is the online sensing-parameter estimation noise. For each sensing path pair, $\tilde{\mathbf{v}}_{(l,l^{\prime}),n}
\sim
\mathcal{N}
(
\mathbf{0},
\tilde{\mathbf{V}}_{(l,l^{\prime}),n}
)$,
with $\tilde{\mathbf{V}}_{(l,l^{\prime}),n}
=
\mathrm{diag}
(
\sigma_{\theta_{l,n}}^{2},
\sigma_{\theta_{l^{\prime},n}}^{2},
\sigma_{\tau_{(l,l^{\prime}),n}}^{2},
\sigma_{\nu_{(l,l^{\prime}),n}}^{2}
)$.
According to (\ref{sigomega}) $\tilde{\mathbf{V}}_{(l,l^{\prime}),n}
=
(
\Gamma_{(l,l^{\prime}),n}^{S}
(\mathbf{w}_n)
)^{-1}
\mathbf{D}$,
where
$\mathbf{D}
=
\mathrm{diag}
\left(
c_{\Omega_1},
c_{\Omega_2},
c_{\Omega_3},
c_{\Omega_4}
\right)$ and
$\tilde{\mathbf{V}}_n
=
\mathrm{blkdiag}
(
\tilde{\mathbf{V}}_{(1,1),n},
\dots,
\tilde{\mathbf{V}}_{(L,L^{\prime}),n}
)$.

In LoS-dominated propagation, the sensing parameters admit a closed-form geometric relation with the node position and velocity. In NLoS propagation, however, the path angles, delays, and Doppler shifts also depend on the surrounding reflectors and scatterers. Hence, without explicit environmental information, a closed-form measurement function solely in terms of the node state is generally unavailable. The NLoS model can therefore be written as
\begin{equation}
\mathbf{z}_n
=
f
\left(
\mathbf{s}_n,
\mathcal{E}
\right)
+
\tilde{\mathbf{v}}_n,
\end{equation}
where $\mathcal{E}$ denotes the propagation environment,  including the
effects of surrounding scatterers and reflectors. In the proposed framework, this environmental dependence is represented through an offline CKM.

Based on \eqref{M_trck} and the node velocity, the complete CKM-assisted sensing-parameter prediction vector, $\mathbf{g}_{\mathrm{CKM}}$, for path pair $(l,l^{\prime})$  is denoted by 
\begin{equation}
\mathbf{g}_{\mathrm{CKM},(l,l^{\prime})}
\left(
\mathbf{s}_n
\right)
=
\begin{bmatrix}
\theta_{l,n}
\\
\theta_{l^{\prime},n}
\\
\tau_{l,n}
+
\tau_{l^{\prime},n}
\\
\left(
\boldsymbol{\kappa}_{l,n}
+
\boldsymbol{\kappa}_{l^{\prime},n}
\right)^{T}
\dot{\mathbf{p}}_n
\end{bmatrix},
\end{equation}
and the aggregate function is
$\mathbf{g}_{\mathrm{CKM}}
\left(
\mathbf{s}_n
\right)
=
[
\mathbf{g}_{\mathrm{CKM},(1,1)}^{T}
\left(
\mathbf{s}_n
\right),
\dots,
\mathbf{g}_{\mathrm{CKM},(L,L^{\prime})}^{T}
\left(
\mathbf{s}_n
\right)
]^{T}$.

According to \eqref{Zn}, $\mathbf{z}_n$ is estimated online from the echo signals received during the current frame. Thus, the CKM does not generate the online observation; it provides its state-dependent prediction for the EKF update.
Accounting for CKM imperfections, the measurement model is written as
\begin{equation}
\mathbf{z}_n
=
\mathbf{g}_{\mathrm{CKM}}
\left(
\mathbf{s}_n
\right)
+
\bar{\mathbf{v}}_{\mathrm{CKM},n}
+
\tilde{\mathbf{v}}_n,
\end{equation}
where $\bar{\mathbf{v}}_{\mathrm{CKM},n}
\sim
\mathcal{N}
\left(
\mathbf{0},
\bar{\mathbf{V}}_{\mathrm{CKM},n}
\right)$
accounts for offline map-construction errors, calibration mismatch, and spatial interpolation. 
Since Doppler is not stored in the CKM, its CKM-induced error results from uncertainty in the stored geometric coefficients. In particular,
$\Delta
\nu_{(l,l^{\prime}),n}
=
\left(
\Delta
\boldsymbol{\kappa}_{l,n}
+
\Delta
\boldsymbol{\kappa}_{l^{\prime},n}
\right)^{T}
\dot{\mathbf{p}}_n$.

The CKM-induced and online sensing-estimation errors arise from distinct offline and online mechanisms and are assumed independent $\mathbb{E}
\left[
\bar{\mathbf{v}}_{\mathrm{CKM},n}
\tilde{\mathbf{v}}_n^{T}
\right]
=
\mathbf{0}$.
Defining $\mathbf{e}_n
=
\bar{\mathbf{v}}_{\mathrm{CKM},n}
+
\tilde{\mathbf{v}}_n$,
the measurement model becomes
\begin{equation}
\mathbf{z}_n
=
\mathbf{g}_{\mathrm{CKM}}
\left(
\mathbf{s}_n
\right)
+
\mathbf{e}_n,
\end{equation}
with effective covariance
\begin{equation}
    \mathbf{V}_n^{\text{eff}}
=
\bar{\mathbf{V}}_{\mathrm{CKM},n}
+
\tilde{\mathbf{V}}_n.
\label{Veff}
\end{equation}

For tractability, $\bar{\mathbf{V}}_{\mathrm{CKM},n}$ is approximated as diagonal, retaining the local marginal uncertainties of the angular, delay, and Doppler components while neglecting cross-parameter and inter-location correlations.

In practice, the number and identities of resolvable paths may vary over time and may not always coincide with those stored in the CKM. The present work assumes a fixed path set over the considered region and time interval, while path birth, disappearance, and online-to-CKM path association are left for future work.
\begin{figure*}[!t]
    \begin{equation}
    \begin{aligned}
        \mathbf{J}_n = \frac{\partial \mathbf{g}_{\text{CKM}}}{\partial \mathbf{s}}|_{\mathbf{s}=\hat{\mathbf{s}}_{n|n-1}} = &
        \Big[ \frac{\mathbf{g}_{\text{CKM}}(\hat{x}_{n} +\Delta \hat{x}_{n},y_{n},\hat{\dot{x}}_{n},\hat{\dot{y}}_{n})-\mathbf{g}_{\mathrm{CKM}}(\mathbf{s})}{\Delta \hat{x}_{n}}, \frac{\mathbf{g}_{\mathrm{CKM}}(\hat{x}_{n}, \hat{y}_{n} +\Delta \hat{y}_{n},\hat{\dot{x}}_{n},\hat{\dot{y}}_{n})-\mathbf{g}_{\mathrm{CKM}}(\mathbf{s})}{\Delta \hat{y}_{n}}, \\
        &\frac{\mathbf{g}_{\mathrm{CKM}}(\hat{x}_{n} ,\hat y_{n},\hat{\dot{x}}_{n}+ \Delta \hat{\dot{x}}_{n},\hat{\dot{y}}_{n})-\mathbf{g}_{\mathrm{CKM}}(\mathbf{s})}{\Delta \hat{\dot{x}}_{n}},  \frac{\mathbf{g}_{\mathrm{CKM}}(\hat x_{n} , \hat y_{n},\hat{\dot{x}}_{n},\hat{\dot{y}}_{n}+ \Delta \hat{\dot{y}}_{n})-\mathbf{g}_{\mathrm{CKM}}(\mathbf{s})}{\Delta \hat{\dot{y}}_{n}} \Big]
        \label{Jakobian}
    \end{aligned}
    \end{equation}
\end{figure*}
\subsection{EKF-Based State Update} The EKF steps are expressed as follows \cite{Kalman1960,CKMTracking}
\begin{itemize}
    \item Step 1: State and Covariance Prediction
    \begin{equation}
        \hat{\mathbf{s}}_{n|n-1} = \mathbf{F} \, \hat{\mathbf{s}}_{n-1},
        \label{StatePred}
    \end{equation}
        \begin{equation}
        \mathbf{\mathbf{\Phi}}_{n|n-1}= \mathbf{F} \, \mathbf{\mathbf{\Phi}}_{n-1} \, \mathbf{F}^T + \mathbf{V}_n.
        \label{CovariancePred}
    \end{equation}

    \item Step 2: Measurement and Innovation Covariance Computation
    \begin{equation}
        \hat{\mathbf{z}}_n = \mathbf{g}_{\text{CKM}}(\hat{\mathbf{s}}_{n|n-1}),
    \end{equation}
    \begin{equation}
        \mathbf{Q}_n = \mathbf{J}_n \mathbf{\mathbf{\Phi}}_{n|n-1} \mathbf{J}_n^T + \mathbf V_{n}^{\text{eff}}
        \label{Q_n}
    \end{equation}
where $\mathbf{J}_n$ is the Jacobian of the measurement model $\mathbf{g}_{\text{CKM}}$ with respect to the state vector $\mathbf{s}$. It is numerically computed as in (\ref{Jakobian}), using small coordinate perturbations $\Delta x_n$ and $\Delta y_n$ to capture the variation of the multipath parameters.

    \item Step 3: Kalman Gain Computation
    \begin{equation}
        \mathbf{K}_n =  \mathbf{\mathbf{\Phi}}_{n|n-1} \, \mathbf{J}_n^T \, (\mathbf{Q}_n)^{-1}.
        \label{KalmanGain}
    \end{equation}

    \item Step 4: State and Covariance Matrix Update
    \begin{equation}
         \hat{\mathbf{s}}_{n} =  \hat{\mathbf{s}}_{n|n-1} + \mathbf{K}_n (\mathbf{z}_n - \hat{\mathbf{z}}_n),
         \label{StateUpdate}
    \end{equation}
    \begin{equation}
        \mathbf{\mathbf{\Phi}}_{n}= (\mathbf{I}_4 - \mathbf{K}_n \, \mathbf{J}_n)  \, \mathbf{\mathbf{\Phi}}_{n|n-1}.
        \label{CovarianceUpd}
    \end{equation}
\end{itemize}
By iteratively updating $\hat{\mathbf{s}}_{n}$ and $\mathbf{\mathbf{\Phi}}_{n}$, the BS tracks the IoT node. The next section presents the real-time beamforming design based on the estimated state $\hat{\mathbf{s}}_{n}$.

\section{Proposed Solution to $P0$}
Problem $\mathrm{P0}$ is non-convex due to its fractional objective and coupled variables. To solve it, we adopt a Dinkelbach-based fractional programming and AO framework~\cite{dinkelbach1967nonlinear}. For a given $\lambda_n \ge 0$, the transformed objective is defined as
\begin{equation}
\mathcal{D}_{\lambda_n}(\mathbf w_n,\rho_n)
\triangleq
\hat P_n^h(\mathbf w_n,\rho_n)
-
\lambda_n P_n^{\text{tot}}(\mathbf w_n).
\label{Dinkelbach_objective}
\end{equation}
Accordingly, for a given $\lambda_n$, problem $\mathrm{P0}$ is transformed into the following subtractive-form problem
\begin{equation}
\begin{aligned}
\mathrm{P1}: \quad \max_{\mathbf{w}_n,\rho_n} \quad 
& \hat{P}_n^{h}(\mathbf{w}_n, \rho_n)
- \lambda_n P_n^{\text{tot}}(\mathbf{w}_n) \\
\textrm{s.t.} \quad 
& \text{C}_1, \text{C}_2, \text{C}_3, \text{C}_4, \text{C}_5.
\end{aligned}
\label{p1}
\end{equation}
For the optimal $\lambda_n^\star$, $\mathrm{P1}$ is equivalent to $\mathrm{P0}$. However, due to the coupling between the beamforming vector and the PS ratio, $\mathrm{P1}$ remains non-convex. Hence, an AO framework is adopted to alternately optimize each variable while fixing the others.
\subsection{Subproblem 1: Transmit Beamforming Optimization}
For a fixed PS ratio $\rho_n$, problem $\mathrm{P1}$ reduces to the following beamforming optimization problem:
\begin{equation}
\begin{aligned}
\mathrm{P2}: \quad \max_{\mathbf{w}_n} \quad 
& \hat{P}_n^{h}(\mathbf{w}_n, \rho_n)
- \lambda_n P_n^{\text{tot}}(\mathbf{w}_n) \\
\textrm{s.t.} \quad 
& \text{C}_1, \text{C}_2, \text{C}_3, \text{C}_4.
\end{aligned}
\label{p2}
\end{equation}
Problem $\mathrm{P2}$ remains non-convex due to the quadratic beamforming terms. We therefore define $\mathbf{W}_n=\mathbf{w}_n\mathbf{w}_n^H$, with $\mathbf{W}_n\succeq \mathbf{0}$ and $\mathrm{rank}(\mathbf{W}_n)=1$. Moreover, since $\mathrm{C}_4$ involves the EKF covariance inversion and is not directly CVX-compatible, we use the equivalent information-filter form \cite[Ch.~6]{simon2006optimal}
\begin{equation}
\mathbf{\Phi}_n=
\left(
\mathbf{\Phi}_{n|n-1}^{-1}
+
\mathbf{J}_n^{T} (\mathbf{V}_{n}^{\text{eff}})^{-1}\mathbf{J}_n
\right)^{-1}.
\label{eq:info_filter}
\end{equation}
Although \eqref{eq:info_filter} removes the posterior covariance inverse, $(\mathbf{V}_n^{\text{eff}})^{-1}$ remains non-affine in $\mathbf{W}_n$ through $\Gamma_{(l,l'),n}^{S}(\mathbf{W}_n)$. To obtain an exact SDP form, we introduce auxiliary variables $\xi_{\Omega_i,(l,l'),n}\ge0$, replacing ${C_4}$ with $\tilde{{C}}_4$--$\tilde{{C}}_6$. Finally, by applying SDR to the rank-one constraint and using LMIs, $\mathrm{P2}$ is reformulated as (\ref{P3}).
\begin{figure*}[!t]
    \begin{equation}
\begin{aligned}
\mathrm{P3}: \quad \max_{\mathbf{W}_n, \, \xi_{\Omega_i,(l,l'),n}, \, \mathcal{A}} \quad 
& \hat{P}_n^{h}(\mathbf{W}_n, \rho_n)
- \lambda_n P_n^{\text{tot}}(\mathbf{W}_n) \\
\textrm{s.t.} \quad 
& \tilde{C}_1:\mathrm{tr}(\mathbf{W}_n) \leq P_{\text{max}}, \\
& \tilde{C}_2: {\Gamma}_n^{C_\text{min}} \leq \hat{\Gamma}_n^C(\mathbf{W}_n) \leq \Gamma_n^{C_\text{max}}, \\
& \tilde{C}_3: P_n^{h_\text{min}} \leq \hat{P}_n^{h}(\mathbf{W}_n, \rho_n) \leq P_n^{h_\text{max}}, \\
& \tilde{C}_4: \begin{bmatrix}
c_{\Omega_i}
+
\bar{\sigma}_{\Omega_i,n}^{2}
\Gamma_{(l,l'),n}^{{S}}(\mathbf{W}_n)
&
\sqrt{c_{\Omega_i}}
\\
\sqrt{c_{\Omega_i}}
&
1-
\bar{\sigma}_{\Omega_i,n}^{2}
\xi_{\Omega_i,(l,l'),n}
\end{bmatrix}
\succeq \mathbf{0}, \\
& \tilde{C}_5:\begin{bmatrix}
\displaystyle
\mathbf{\Phi}_{n|n-1}^{-1}
+
\sum_{l=1}^{L}
\sum_{l'=1}^{L'}
\mathbf{J}_{(l,l'),n}^{T}
\operatorname{diag}
\left(
\xi_{\Omega_i,(l,l'),n}
\right)
\mathbf{J}_{(l,l'),n}
&
\mathbf{I}_4
\\
\mathbf{I}_4
&
\mathcal{A}_n
\end{bmatrix}
\succeq \mathbf{0},\\
& \tilde{C}_6: \mathrm{tr}(\bar{\mathcal{A}}_n)\le \mathrm{tr}(\bar{\mathbf{\Phi}})_{\text{max}},\\
& \tilde{C}_7: \mathbf{W}_n \succeq \mathbf{0}.
\end{aligned}
\label{P3}
\end{equation}
\vspace{1mm}
\noindent\rule{\linewidth}{0.4pt}
\end{figure*}
Here, $\hat{\Gamma}_n^{C}=2^{\hat{R}_n}-1$, $\mathbf{J}_{(l,l'),n}\in\mathbb{R}^{4\times4}$ is the Jacobian block for path pair $(l,l')$, and $\mathcal{A}_n\in\mathbb{S}^{4}$ is an auxiliary matrix. The derivation of $\tilde{C}_4$--$\tilde{C}_6$ is provided in the Appendix. For $\bar{\sigma}_{\Omega_i,n}^{2}=0$, the corresponding LMI reduces to $c_{\Omega_i}\xi_{\Omega_i,(l,l'),n}\leq\Gamma_{(l,l'),n}^{S}(\mathbf{W}_n)$. Moreover,
$\mathrm{tr}(\mathcal{\bar{A}})=\mathrm{tr}\left(\mathcal{T}^{-1}\mathcal{A}_n\mathcal{T}^{-T}\right)$.
Problem $\mathrm{P3}$ is convex and can be efficiently solved using CVX, yielding the optimal solution $\mathbf{W}_n^{\star}$.
 
\textbf{Proposition 1} :
For any fixed PS ratio $\rho_n$ and Dinkelbach parameter $\lambda_n$, consider the semidefinite program $\mathrm{P3}$ in (\ref{P3}). Assume that $\mathrm{P3}$ is feasible and admits a nonzero optimal transmit covariance matrix $\mathbf W_n^\star\succeq\mathbf 0$. Furthermore, assume that the BS employs a ULA and that the objective function and all constraints of $\mathrm{P3}$ depend on $\mathbf W_n$ only through $\operatorname{tr}(\mathbf W_n)$ and the directional transmit-power terms $\mathbf a_T^H(\theta_l)\mathbf W_n\mathbf a_T(\theta_l),
\qquad l=1,\ldots,L$.

The SDR in $\mathrm{P3}$ is tight, i.e., there always exists an optimal rank-one solution $\bar{\mathbf W}_n$. If $\mathbf W_n^\star$ is rank-one, the beamforming vector is directly recovered via eigenvalue decomposition (EVD). Otherwise, the high-rank solution only reflects non-uniqueness; by Proposition 1, an equivalent rank-one optimum exists.

\subsection{Subproblem 2: Power Splitting Optimization}
For a fixed beamforming vector $\mathbf{w}$, problem $\mathrm{P1}$ reduces to the PS optimization problem
\begin{equation}
\begin{aligned}
\mathrm{P4}: \quad \max_{\rho_n} \quad 
& \hat{P}_n^{h}(\mathbf{w}_n, \rho_n)
- \lambda_n P_n^{\text{tot}}(\mathbf{w}_n) \\
\textrm{s.t.} \quad 
& \text{C}_2, \text{C}_3, \text{C}_5.
\end{aligned}
\label{p4}
\end{equation}
Since $P_n^{\text{tot}}(\mathbf{w}_n)$ is independent of $\rho_n$, $\mathrm{P4}$ reduces to maximizing $E_n^{h}$. Defining $\mathcal{B} \triangleq \sum_{l=1}^L  |\alpha_{l}|^2 \mathbf{a}_T^H(\theta_{l}) \mathbf{w}_{n} \mathbf{w}_{n}^H \mathbf{a}_T(\theta_{l})$, the feasible range of $\rho_n$ follows from the rate and harvested-energy constraints, yielding the reformulated problem
\begin{equation}
\begin{aligned}
\mathrm{P5}: \quad \max_{\rho_n} \quad 
& \hat{P}_n^{h}(\mathbf{w}_n, \rho_n) \\
\textrm{s.t.} \quad 
& \bar{\text{C}_2}: \frac{\Gamma_n^{C_\text{min}} \,  \tilde{\delta}^2}{\mathcal{B}- \Gamma_n^{C_\text{min}} \, \delta^2} \leq \rho_n \leq \frac{\Gamma_n^{C_\text{max}}\,  \tilde{\delta}^2}{\mathcal{B}- \Gamma_n^{C_\text{max}} \, \delta^2}   ,\\
&\bar{\text{C}_3}, 1- \frac{P_n^{h_\text{max}}}{\zeta_{n}(\mathcal{B}+\delta^2)} \leq \rho_n \leq 1- \frac{P_n^{h_\text{min}}}{\zeta_{n}(\mathcal{B}+\delta^2)} \\
&\bar{\text{C}_5}: 0 < \rho_n < 1.
\end{aligned}
\end{equation}

Since $P_n^{\text{tot}}(\mathbf{w}_n)$ is independent of $\rho_n$, $\mathrm{P4}$ is feasible iff $\rho_n^{\mathrm{L}}\leq\rho_n^{\mathrm{U}}$, with the optimal solution $\rho_n^\star=\rho_n^{\mathrm{L}}$.

The Dinkelbach-based AO procedure is summarized in Algorithm~1. Building on it, the proposed CKM-based predictive beamforming for mobile tracking (CKM-PBMT) combines CKM-based channel prediction and mobility tracking, while Algorithm~1 optimizes the beamforming variables at each frame. The overall CKM-PBMT procedure is summarized in Algorithm~2.

\section{Simulation Results}
In this section, we evaluate the proposed CKM-PBMT framework for a $28~\mathrm{GHz}$ ISCAP downlink system. The BS employs ULA with $N_t=16$ and $N_r=8$ antennas to serve a single-antenna IoT node with a power splitter and finite-capacity battery. Simulations are conducted over a $30\times20~\mathrm{m}^2$ area, with the BS at $(3,10)~\mathrm{m}$ and the node initialized at $(7.5,7)~\mathrm{m}$ with velocity $(1.4,0.3)~\mathrm{m/s}$. A CV model with $T_f=15~\mathrm{ms}$ is assumed.

The CKM covers a rectangular region with $x\in[4.5,18]~\mathrm{m}$ and $y\in[4,13]~\mathrm{m}$, using a spatial resolution of $0.1~\mathrm{m}$. It includes three NLoS paths, resulting in nine sensing path pairs. CKM errors are modeled as spatially correlated random fields with standard deviations of $1~\mathrm{dB}$, $0.5^\circ$, $ 1~\mathrm {ns}$, and $0.25~\mathrm{Hz/(m/s)}$ for path gain, angle, delay, and Doppler coefficient, respectively.

We set $P_{\max}=38~\mathrm{dBm}$, PA efficiency to $0.4$, transmitter/receiver circuit powers to $2~\mathrm{W}$/$10^{-4}~\mathrm{W}$, energy-conversion efficiency to $0.8$, minimum harvested power to $10^{-8}~\mathrm{W}$, and decoding coefficient to $10^{-4}~\mathrm{W/(bit/s/Hz)}$. The battery capacity and initial energy are $1.5~\mathrm{mJ}$ and $1.1~\mathrm{mJ}$. The minimum C-SNR is $5~\mathrm{dB}$, increased to $15~\mathrm{dB}$ for convergence tests, and the sensing gain is $40~\mathrm{dB}$. The nominal horizon is $50$ frames and $70$ frames for battery management.

For fairness, all methods use identical trajectories, process noise, initial EKF errors, and CKM-error realizations; only the online sensing noise varies across Monte Carlo runs.

\subsection{Validation of the CKM Structure and Accuracy}
The CKM is constructed offline using a geometry-based model of a two-dimensional indoor factory hall containing the BS, boundary walls, a metallic obstacle, and an internal partition. The node mobility region is discretized into a spatial grid, where the dominant single-bounce reflected paths are computed and stored. A path is retained only if its reflection point lies on the corresponding physical reflector and satisfies the geometric validity conditions.

To account for CKM imperfections, two map versions are considered. The ground-truth map, CKM$_{\text{true}}$, generates the actual channel, sensing measurements, QoS, and energy consumption, whereas CKM$_{\text{est}}$, used by the proposed beamforming and EKF algorithms, is obtained by perturbing CKM$_{\text{true}}$. The perturbations affect the path gains, angles, delays, and Doppler coefficients and are modeled as spatially correlated random fields to capture map-construction, calibration, measurement, and ray-tracing errors. Hence, the simulations account for both CKM uncertainty and state-prediction errors while preserving the spatial consistency of the map.
\begin{figure*}[!t]
    \centering

    \subfloat[!t]{%
        \includegraphics[width=0.31\textwidth]{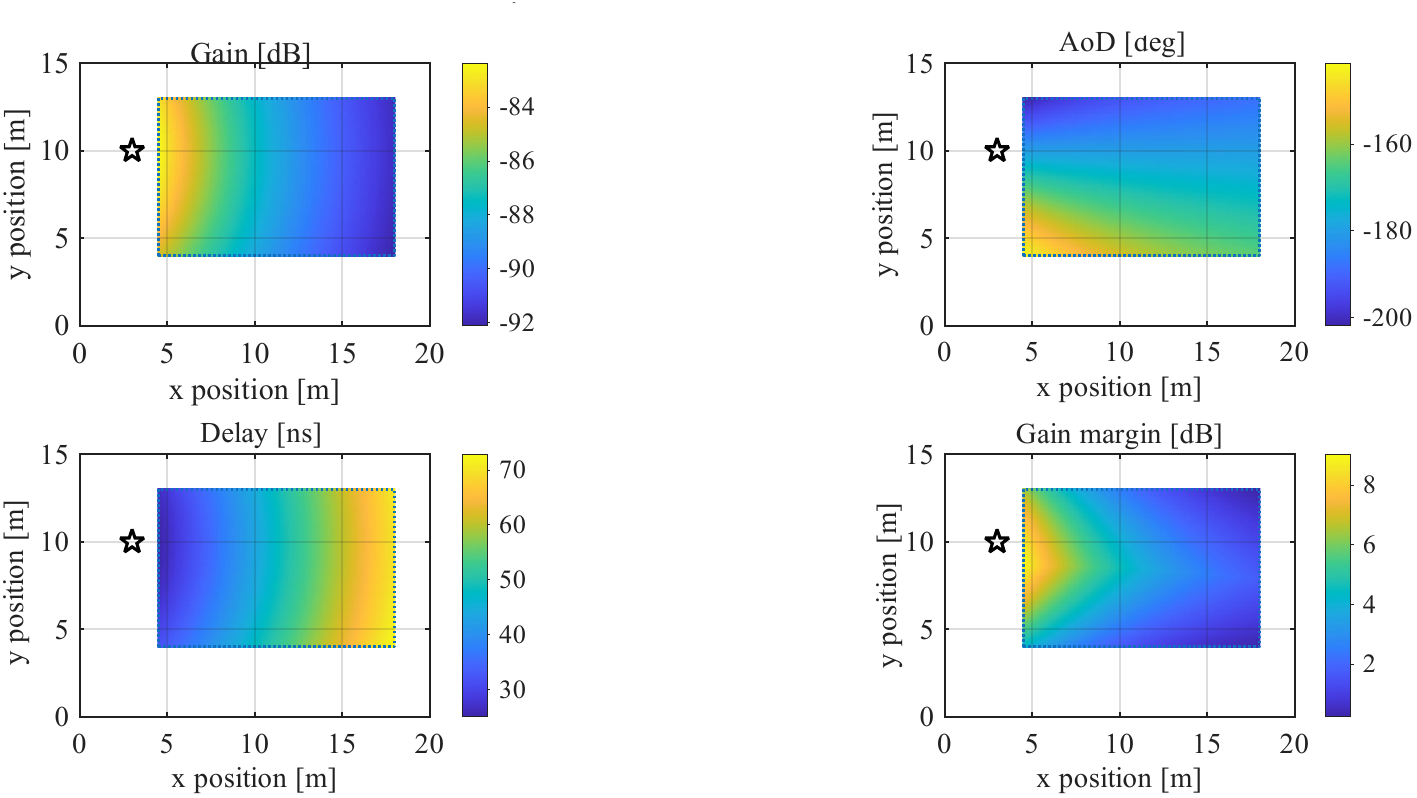}%
        \label{fig:ckm_maps}
    }
    \hfill
    \subfloat[!t]{%
        \includegraphics[width=0.31\textwidth]{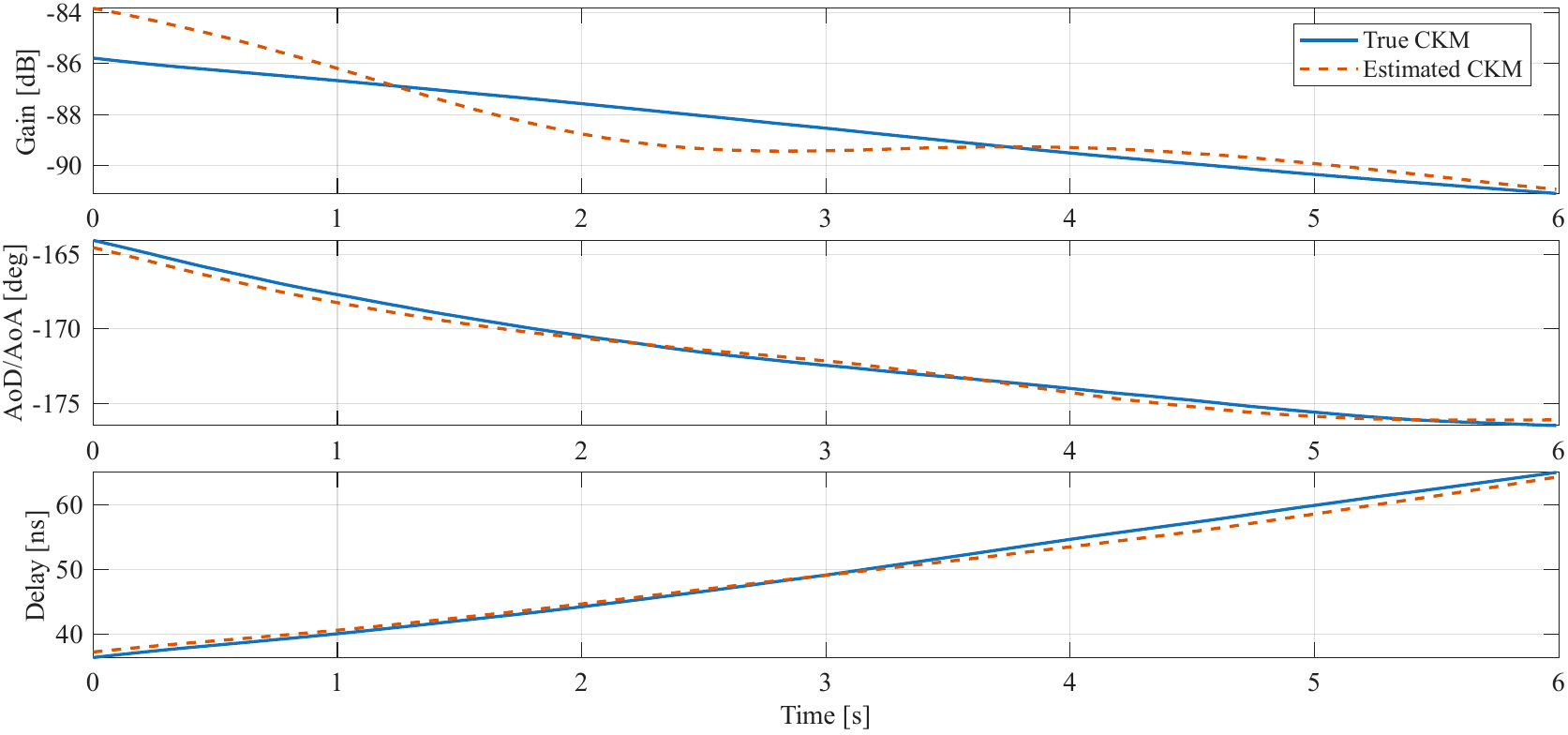}%
        \label{fig:true_est_trajectory}
    }
    \hfill
    \subfloat[!t]{%
        \includegraphics[width=0.31\textwidth]{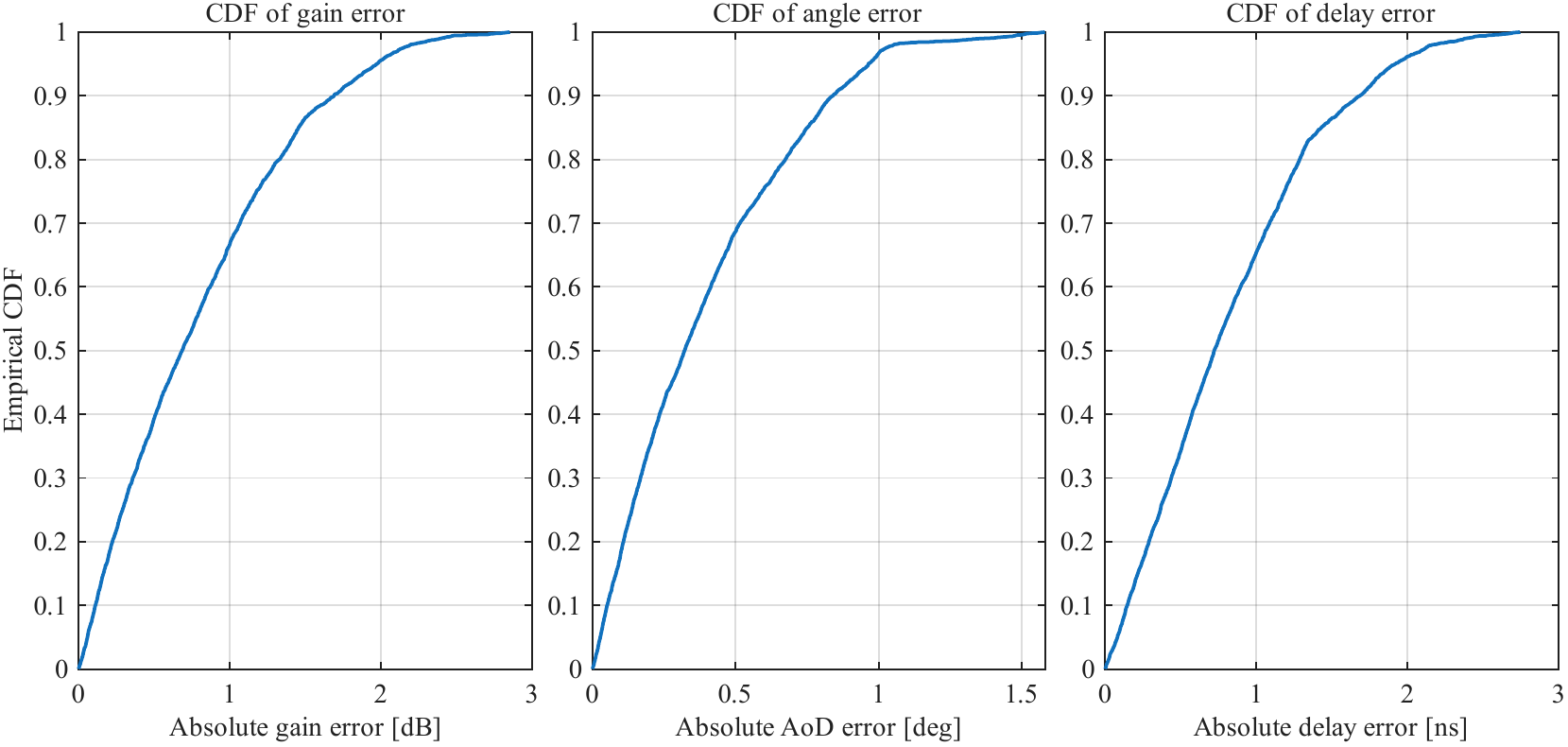}%
        \label{fig:ckm_error_cdf}
    }

    \caption{CKM validation results: (a) Ground-truth spatial structure of the dominant NLoS CKM path, where the black star indicates the BS location; (b) true CKM versus estimated CKM along the trajectory; and (c) CKM error CDFs.}
    \label{fig:ckm_validation}
\end{figure*}

Fig.~\ref{fig:ckm_maps} depicts the spatial distributions of the dominant-path gain, AoD, delay, and gain margin between the two strongest paths. Their smooth variations across the CKM support confirm the spatial consistency of the generated channel parameters. In general, the dominant-path gain decreases with effective propagation distance, while delay increases with path length. The AoD also varies gradually with location, and the gain-margin map identifies regions where the dominant path is more clearly separated from the others.

Fig.~\ref{fig:true_est_trajectory} compares the selected-path parameters along the node trajectory with their ground-truth values in CKM$_{\text{true}}$. The estimated map, CKM$_{\text{est}}$, captures the main variations in gain, angle, and delay with limited deviations consistent with the adopted perturbation models. The gain is slightly overestimated in some regions, the AoD mismatch remains within the prescribed angular-error range, and the delay error is comparable to the applied timing perturbation. Thus, despite being imperfect, CKM$_{\text{est}}$ preserves the underlying spatial trends and provides useful location-dependent information for prediction, state estimation, beamforming, and resource allocation.

Fig.~\ref{fig:ckm_error_cdf} presents the CDFs of gain, AoD, and delay errors over test points within the CKM support. Most gain and angular errors remain within a few decibels and degrees, respectively, while delay errors are generally limited to a few nanoseconds. The absence of excessive tails is important because low average error or RMSE can still conceal severe local errors. The CDF results therefore confirm that CKM$_{\text{est}}$ is accurate on average while maintaining controlled errors across the spatial domain.

\subsection{Convergence Analysis of Algorithm 1}

\begin{figure}[!t]
    \centering
    \includegraphics[width=1.01\linewidth]{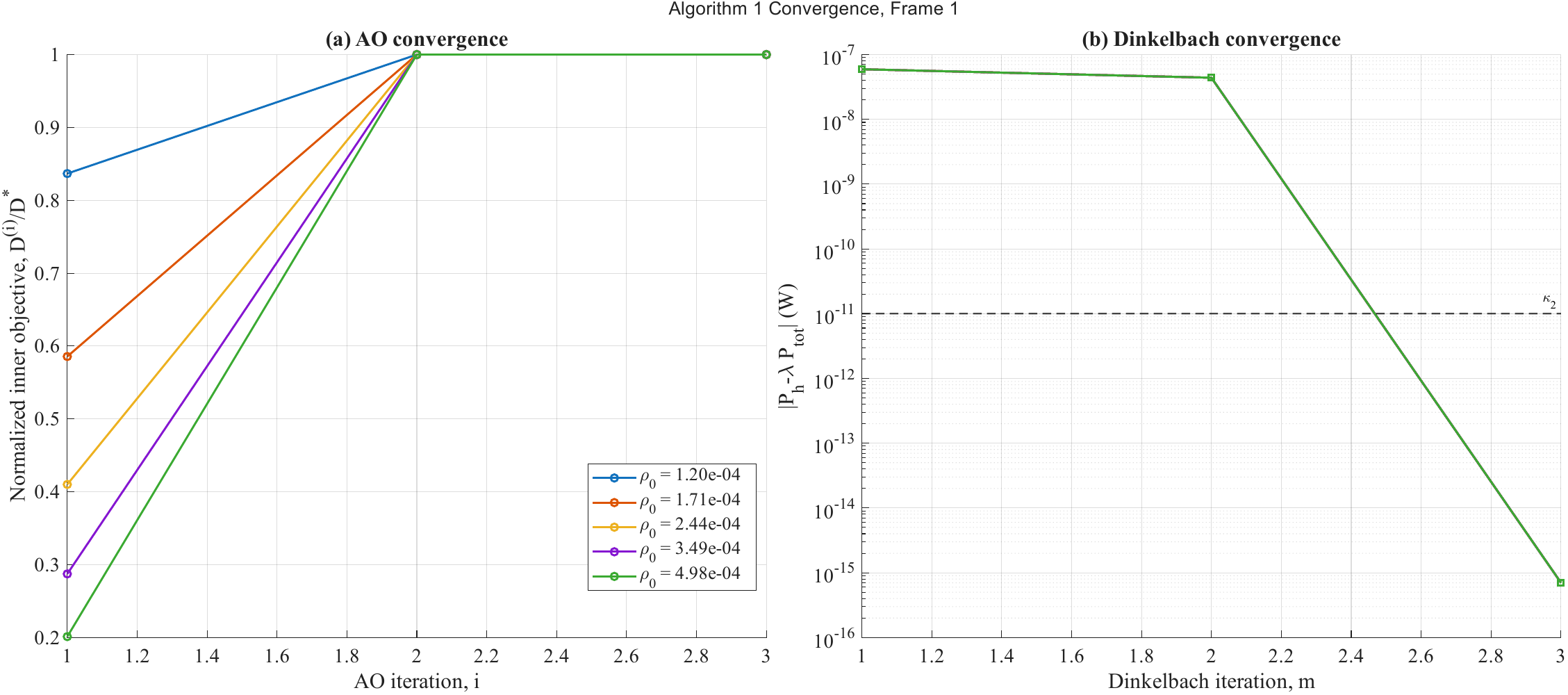}
    \caption{Convergence Behavior of Algorithm 1.}
    \label{fig:Alg1}
\end{figure}

\begin{figure*}[!t]
    \centering
    \subfloat[!t]{%
        \includegraphics[width=0.31\textwidth]{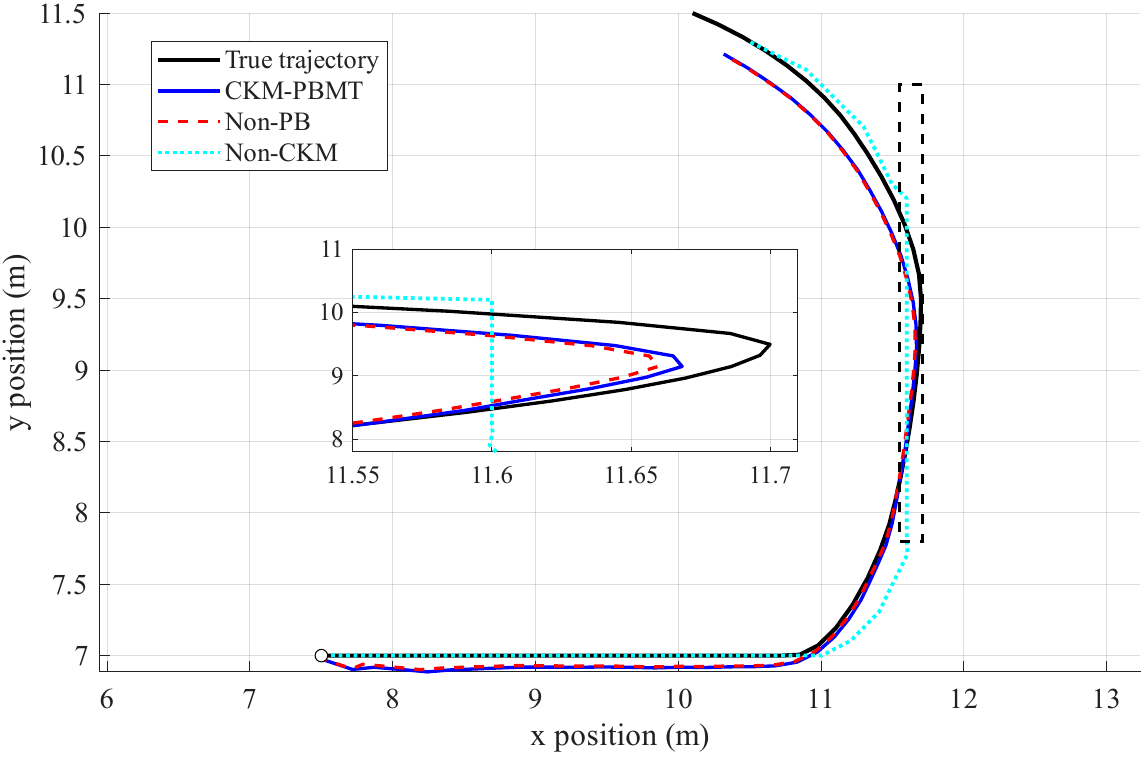}%
        \label{fig:Trck1}
    }
    \hfill
    \subfloat[!t]{%
        \includegraphics[width=0.31\textwidth]{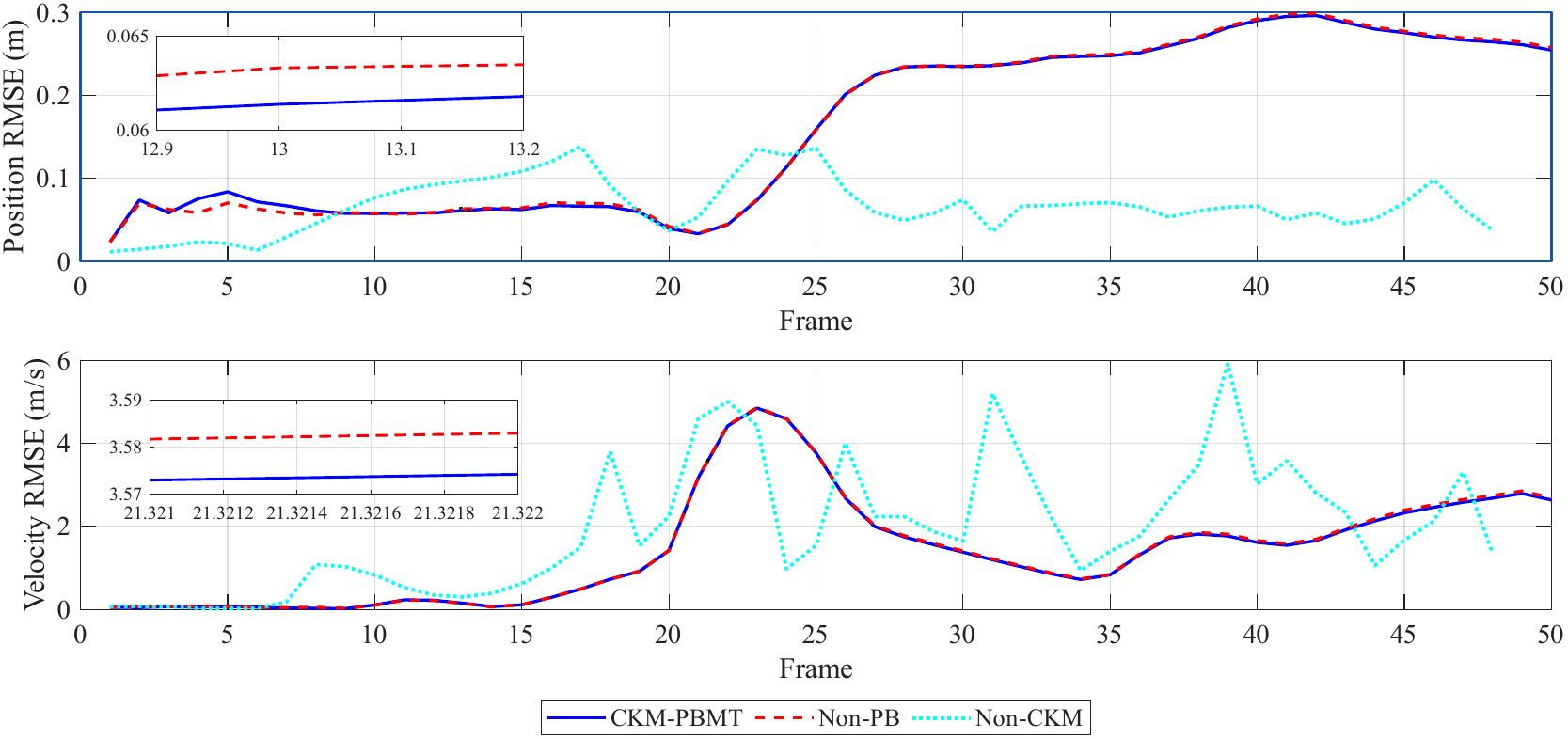}%
        \label{fig:Trck2}
    }
    \hfill
    \subfloat[!t]{%
        \includegraphics[width=0.31\textwidth]{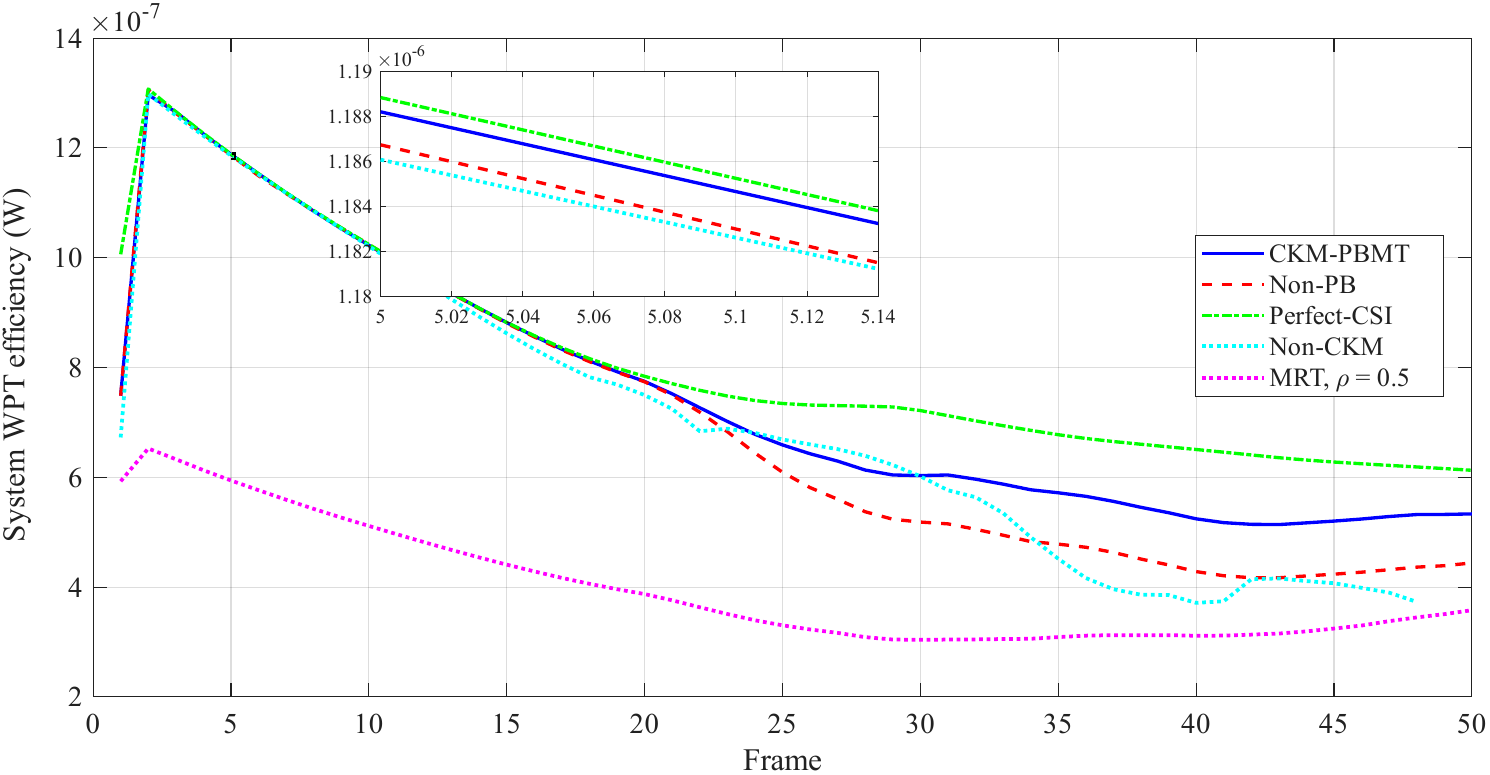}%
        \label{fig:Trck3}
    }
    \caption{Performance Evaluation of the CKM-PBMT Framework: (a)  representative true and EKF-estimated trajectories, (b) position and velocity tracking RMSE, and (c) WPT efficiency versus Frame.}
    \label{fig:Track}
\end{figure*}
Fig.~\ref{fig:Alg1} illustrates the convergence behavior of Algorithm~1 for the AO and Dinkelbach iterations. To evaluate the sensitivity of the AO procedure to initialization, several distinct feasible cold-start values of the power-splitting ratio $\rho_0$ are considered. As shown in Fig.~\ref{fig:Alg1}a, although the initial objective values differ considerably, all trajectories converge to the same normalized objective value within two AO iterations, indicating low sensitivity to the tested initializations. This rapid convergence is consistent with the structure of the proposed AO procedure, where the beamforming and power-splitting variables are optimized alternately through the corresponding subproblems.

Fig.~\ref{fig:Alg1}b shows the convergence of the outer Dinkelbach loop, where the residual $| P_n^{h} - \lambda_n P_n^{\text{tot}}|$
decreases and falls below the prescribed tolerance $\kappa_2$ within three iterations. These results demonstrate the fast and numerically stable convergence of Algorithm~1 while showing that the observed convergence behavior is robust to different feasible initializations.

\subsection{Performance Evaluation of the CKM-PBMT Framework}
The proposed CKM-PBMT is compared with four baseline schemes: Non-Predictive Beamforming (Non-PB), which designs the beam using the previous posterior state instead of the predicted current state; Perfect-CSI, which assumes ideal instantaneous channel knowledge and provides an upper-bound performance reference; MRT, which applies maximum-ratio transmission with a fixed PS ratio of $\rho=0.5$ without performing the proposed joint AO/Dinkelbach optimization; and Non-CKM, which estimates the channel using conventional uplink pilots, performs CSI-fingerprint-based localization, and applies a delayed constant-velocity KF without CKM information.

Fig.~\ref{fig:Trck1} presents the true and estimated trajectories, while Fig.~\ref{fig:Trck2} shows the position and velocity RMSE results. For the relatively unobstructed and nearly straight segment of the trajectory, all schemes achieve comparable tracking performance, resulting in similar WPT performance. The advantage of CKM-PBMT becomes more apparent when the IoT node approaches and moves around the obstacle, where the propagation geometry and multipath characteristics vary significantly. In this region, CKM-PBMT achieves more accurate tracking and more stable beam alignment than Non-PB by exploiting the predicted state and the location-dependent CKM.

The Non-CKM scheme exhibits larger frame-to-frame variations, particularly in velocity RMSE. This behavior is mainly caused by the noisy CSI-fingerprint localization and the delayed KF. Small variations in the selected fingerprints can affect the estimated position, and these errors are further amplified in the velocity estimation over short frame intervals. Similar fluctuations can be observed in Fig.~\ref{fig:Trck3} in terms of WPT efficiency. Since beamforming relies on CSI-based localization followed by beam design, occasional fingerprint estimation errors may temporarily result in better channel alignment than CKM-PBMT, leading to higher WPT efficiency in some frames. However, the Non-CKM scheme provides less stable performance and requires additional communication and processing overhead due to pilot-based channel acquisition. In the considered simulation, the IoT node using the Non-CKM scheme depletes its battery at frame 49 and the operation terminates. In contrast, CKM enables an environment-aware and location-dependent channel representation, reducing the need for repeated online channel reconstruction and improving the robustness of beam design against instantaneous pilot-estimation errors. This improvement is achieved at the cost of requiring an accurate site-specific CKM.

The MRT scheme achieves lower WPT efficiency because it does not perform the proposed joint optimization. Specifically, the tracking-error constraint is not considered during beam design; therefore, the transmit beam is not adapted to the available tracking accuracy and state uncertainty. This limitation becomes more significant as the propagation geometry changes along the trajectory. The Perfect-CSI benchmark provides the upper-bound reference and illustrates the performance gap caused by imperfect channel knowledge.

The results demonstrate that the primary advantage of CKM-PBMT is not only improved tracking performance in unobstructed regions, where most schemes perform similarly, but also its capability to jointly utilize state prediction, environment-aware CKM, and tracking-aware beamforming. This joint design maintains beam alignment and WPT efficiency in challenging obstacle-induced and multipath-dominated environments.

\subsection{Battery-Aware Communication and Energy Management}
\begin{figure*}[!t]
    \centering

    \subfloat[!t]{%
        \includegraphics[width=0.45\textwidth]{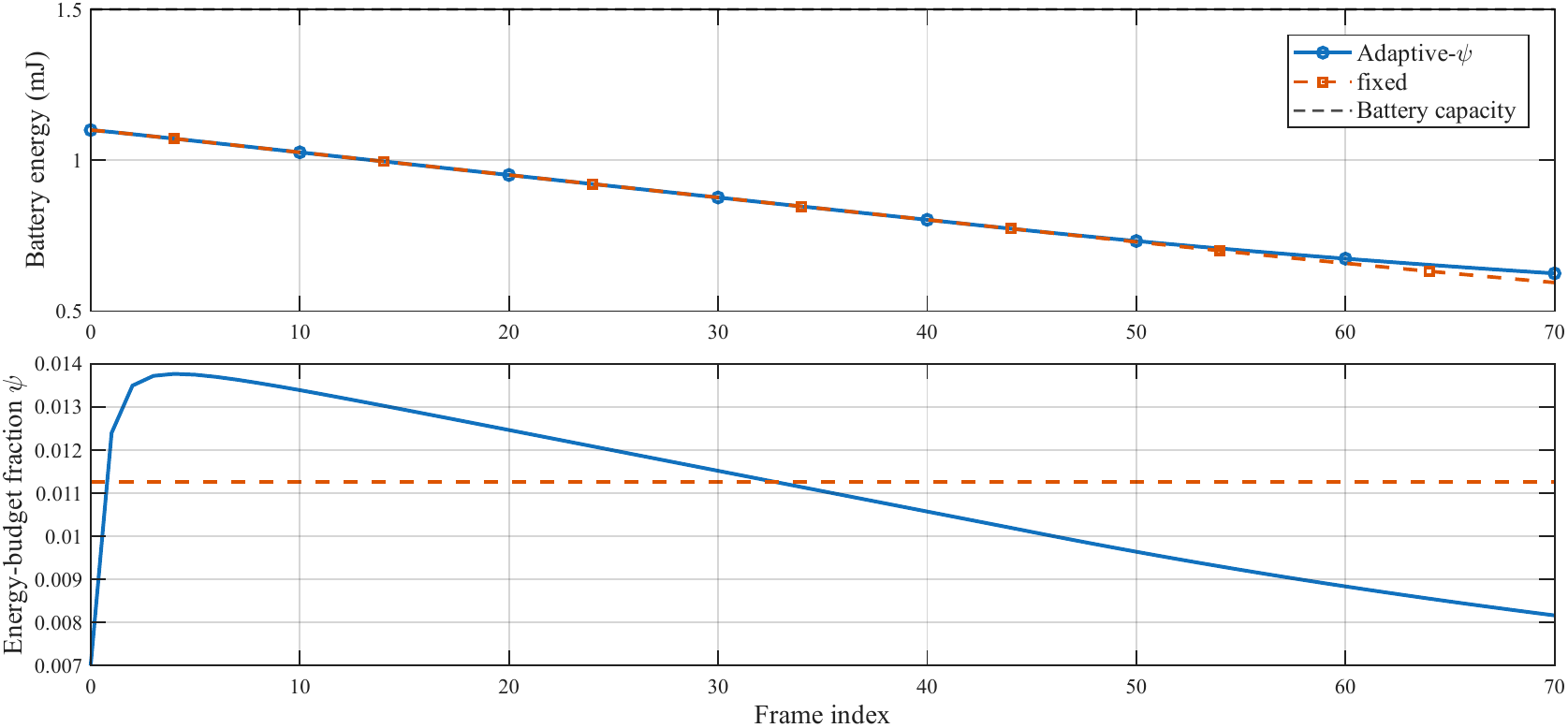}%
        \label{fig:battery_psi}
    }
    \hfill
    \subfloat[!t]{%
        \includegraphics[width=0.45\textwidth]{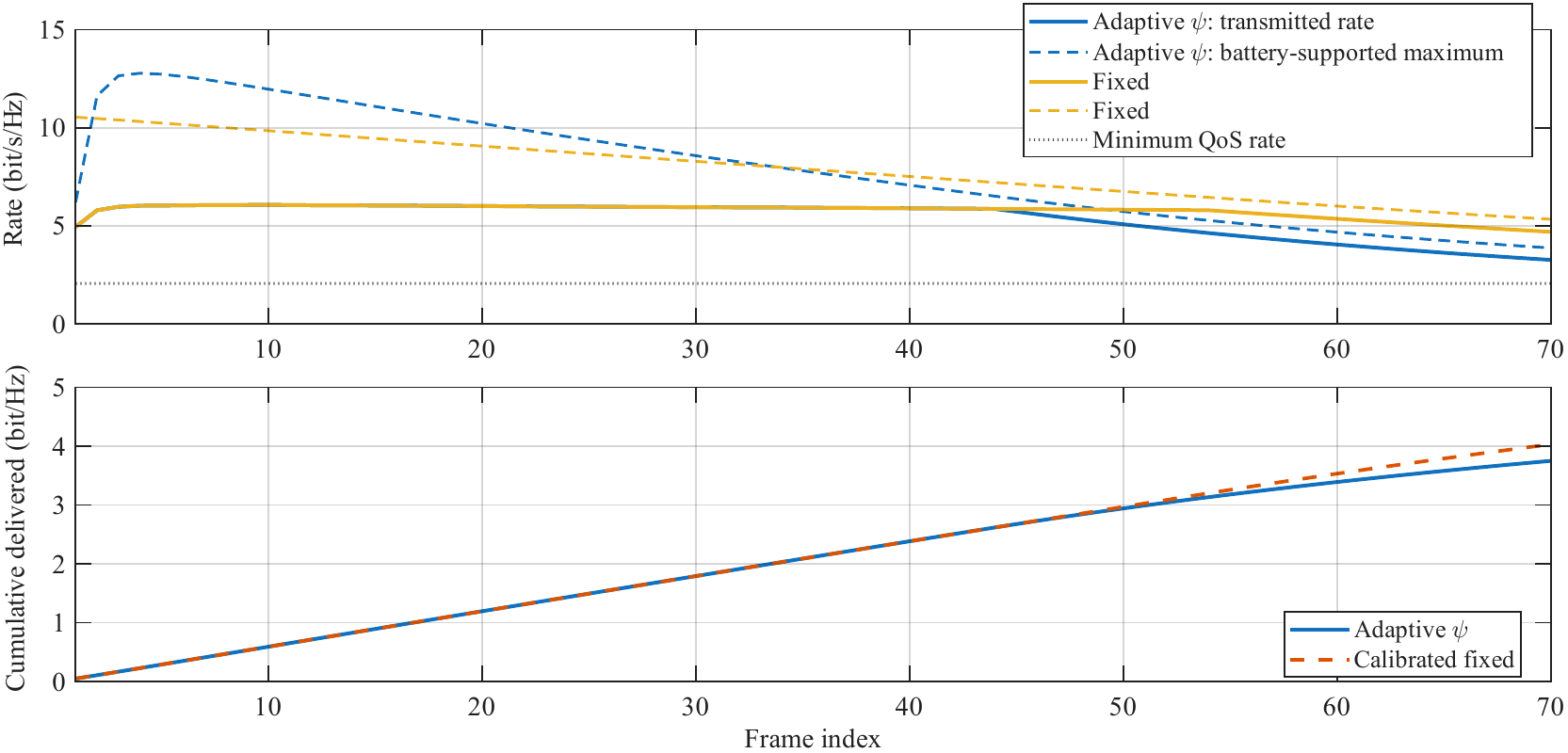}%
        \label{fig:rate_information}
    }

    \caption{Battery and information results: (a) battery evolution under adaptive and fixed energy-budget policies, and (b) communication performance under adaptive and fixed energy-budget policies.}
    \label{fig:battery_information}
\end{figure*}

Fig.~\ref{fig:battery_information} compares the adaptive $\psi_n$ policy with the calibrated fixed policy. As shown in Fig.~\ref{fig:battery_psi}, the adaptive policy allocates more energy to processing and decoding during the initial frames, when the battery level is high, and gradually reduces $\psi_n$ as the stored energy decreases to preserve battery capacity. In contrast, the fixed policy maintains $\psi=0.0113$ regardless of the battery state. Thus, the energy trajectories remain nearly identical during the middle of the horizon, while in the final frames, the fixed policy depletes the battery more rapidly and the adaptive policy retains more residual energy.

As shown in Fig.~\ref{fig:rate_information}, both policies initially achieve nearly the same transmission rate, around $6$~bit/s/Hz. As the battery becomes more constrained, the adaptive policy gradually reduces the rate, whereas the fixed policy maintains a higher rate in the final frames. Consequently, the fixed policy delivers more cumulative information by the end of the horizon, at the cost of greater battery consumption. Nevertheless, both policies satisfy the minimum QoS requirement throughout the horizon. Hence, the main advantage of the Adaptive-$\psi$ policy is not higher absolute throughput, but a dynamic trade-off between communication rate and battery preservation, avoiding rapid energy depletion while maintaining reliable communication performance.
\section{Conclusion}
This paper developed a CKM-based predictive beamforming framework for ISCAP systems with a mobile energy-constrained node operating in a multipath NLoS environment. The proposed design exploits location-dependent channel knowledge for predictive transmission and EKF-based tracking, while jointly optimizing the transmit covariance and power-splitting ratio under communication, sensing, power, and tracking requirements. A position-uncertainty constraint was incorporated to explicitly control tracking quality, and a battery-aware inter-frame policy was introduced to adapt the decoding-energy budget according to the tracking uncertainty and remaining battery energy. The resulting non-convex problem was addressed using Dinkelbach's method, alternating optimization, and semidefinite relaxation. Numerical results showed that the proposed scheme maintains reliable tracking, enforces the prescribed position-uncertainty requirement, approaches the oracle WPT performance after the initial transient, and improves battery preservation while satisfying the communication QoS requirement. These results highlight the potential of CKM-assisted predictive operation for jointly supporting sensing, communication, wireless powering, and energy management in mobile IoT systems.
\begin{algorithm}[!t]
\caption{Dinkelbach-Based AO Beamforming Optimization}
\begin{algorithmic}[1]

\STATE \textbf{Initialize:}
PS ratio $\rho_n^{(0)}$, Dinkelbach parameter
$\lambda_n^{(0)}$, tolerance thresholds $\kappa_1$ and
$\kappa_2$, and iteration indices $m=0$, $i=0$.

\REPEAT

    \STATE Initialize the AO iteration index $i=0$.

    \REPEAT

        \STATE Solve problem $(\mathrm{P3})$ for given
        $\rho_n^{(i)}$ and $\lambda_n^{(m)}$, and obtain
        $\mathbf{W}_n^{(i+1)}$.

        \STATE Recover the beamforming vector
        $\mathbf{w}_n^{(i+1)}$ from
        $\mathbf{W}_n^{(i+1)}$ using EVD or Fejér–Riesz-based deterministic rank-one recovery.

        \STATE Update the PS ratio $\rho_n^{(i+1)}$ by
        solving problem $(\mathrm{P5})$ for the given
        $\mathbf{w}_n^{(i+1)}$.

        \STATE Compute the achieved WPT efficiency as
        \[
        \eta_n^{(i+1)}
        =
        \frac{
        \hat{P}_n^{\mathrm{h}}
        \left(
        \mathbf{w}_n^{(i+1)},
        \rho_n^{(i+1)}
        \right)
        }{
        P_n^{\mathrm{tot}}
        \left(
        \mathbf{w}_n^{(i+1)}
        \right)
        }.
        \]
        \STATE Evaluate the inner-loop subtractive objective as
        \[ D_n^{(i+1)}
=
\hat{P}_n^h
\left(
\mathbf{w}_n^{(i+1)},\rho_n^{(i+1)}
\right)
-
\lambda_n^{(m)}
P_n^{\mathrm{tot}}
\left(
\mathbf{w}_n^{(i+1)}
\right).\]

        \STATE Update $i=i+1$.

    \UNTIL{
    $i\geq 2$ and
    $\left|D_n^{(i)}-D_n^{(i-1)}\right|
    \leq\kappa_1$
    }

    \STATE Set
    \[
    \mathbf{w}_n^{\star}
    =
    \mathbf{w}_n^{(i)},
    \qquad
    \rho_n^{\star}
    =
    \rho_n^{(i)}.
    \]

    \STATE Update the Dinkelbach parameter as
    \[
    \lambda_n^{(m+1)}
    =
    \frac{
    \hat{P}_n^{h}
    \left(
    \mathbf{w}_n^{\star},
    \rho_n^{\star}
    \right)
    }{
    P_n^{\text{tot}}}
    \left(
    \mathbf{w}_n^{\star}
    \right).
    \]

    \STATE Set $\rho_n^{(0)}=\rho_n^{\star}$ and update
    $m=m+1$.

\UNTIL{
\[
\left|
\hat{P}_n^{h}
\left(
\mathbf{w}_n^{\star},
\rho_n^{\star}
\right)
-
\lambda_n^{(m-1)}
P_n^{\text{tot}}
\left(
\mathbf{w}_n^{\star}
\right)
\right|
\leq
\kappa_2
\]
}

\STATE \textbf{Output:}
$\mathbf{w}_n^{\star}$ and $\rho_n^{\star}$.

\label{alg1}

\end{algorithmic}
\end{algorithm}
\begin{algorithm}[!t]
\caption{CKM-PBMT Framework}
\label{alg}
\begin{algorithmic}[1]
\STATE \textbf{Initialize:} $\hat{\mathbf{s}}_{0}$, $\boldsymbol{\mathbf{\Phi}}_{0}$, $E_{0}$, $\psi_0$, $E_{\text{max}}$, $T_{f}$, and the system parameters.
\FOR{$n=1,2,\ldots$}
\STATE Compute harvested power upper bound and maximum allowable rate according to (\ref{Emax}) and (\ref{data-rate}), respectively.
\STATE Predict the node state and covariance according to (\ref{StatePred}),(\ref{CovariancePred}), yielding $\hat{\mathbf{s}}_{n|n-1}$ and $\boldsymbol{\mathbf{\Phi}}_{n|n-1}$.
\STATE Query and cache the CKM output at $\hat{\mathbf{s}}_{n|n-1}$, and extract $\mathcal{M}_{\mathrm{ch}}(\hat{\mathbf{s}}_{n|n-1})$ for channel prediction.
\STATE Compute the Jacobian matrix $\mathbf{J}_{n}$ according to (\ref{Jakobian}).
\STATE Run Algorithm~1 using the predicted C-SNR, S-SNR, and harvested power expressions to obtain the optimal beamforming vector, $\mathbf{w}_{n}^{\star}$, and PS ratio, $\rho_{n}^{\star}$.
\STATE Perform ISCAP signal transmission using $\mathbf{w}_{n}^{\star}$ and obtain the sensing measurement $\mathbf{z}_{n}$ from the received echoes using signal processing methods .
\STATE Retrieve the cached sensing-parameter subset and compute the predicted measurement vector as $\hat{\mathbf z}_n =\mathbf{g}_{\mathrm{CKM}} \left(\hat{\mathbf{s}}_{n|n-1}\right).$
\STATE Compute the sensing-parameter noise covariance matrix $\mathbf{V}_{n}^{\text{eff}}$ from (\ref{Veff}), and the innovation covariance matrix $\mathbf{Q}_{n}$ using(\ref{Q_n}).
\STATE Compute the Kalman gain $\mathbf{K}_{n}$ according to (\ref{KalmanGain}).
\STATE Update the state estimate and covariance matrix using (\ref{StateUpdate}) and (\ref{CovarianceUpd}).
\STATE Update the battery-level and adaptive-energy allocation coefficient $\psi_{n+1}$ according to (\ref{Ctilt}) and (\ref{adaptive_budget_factor}), respectively.
\STATE Compute the allowable energy-consumption budget for the next frame using (\ref{adaptive_energy_budget}).
\ENDFOR 
\end{algorithmic}
\end{algorithm}

\appendices
\section{Proof of the SDP Reformulation of $\mathrm{C4}$}
For each path pair $(l,l')$, define $\gamma_{(l,l'),n}\triangleq\Gamma_{(l,l'),n}^{S}(\mathbf{W}_n)$. From the sensing-error model in \eqref{sigomega} and the effective observation covariance, the inverse covariance block is
\begin{equation}
\left(
\mathbf{V}_{(l,l'),n}^{\text{eff}}(\mathbf{W}_n)
\right)^{-1}=
\operatorname{diag}
\left(
q_{\Omega_1,(l,l'),n},
\ldots,
q_{\Omega_4,(l,l'),n}
\right),
\label{eq:app_inverse_noise}
\end{equation}
where
\begin{equation}
q_{\Omega_i,(l,l'),n}=
\frac{
\gamma_{(l,l'),n}
}{
c_{\Omega_i}
+
\bar{\sigma}_{\Omega_i,n}^{2}
\gamma_{(l,l'),n}
}.
\label{eq:app_information_entry}
\end{equation}
Although $\gamma_{(l,l'),n}$ is affine in $\mathbf{W}_n$, each entry in \eqref{eq:app_information_entry} is fractional and therefore cannot be directly incorporated into an affine LMI.

Introduce the auxiliary variable $\xi_{\Omega_i,(l,l'),n}\geq0$ such that $\xi_{\Omega_i,(l,l'),n}\leq q_{\Omega_i,(l,l'),n}$. For $\bar{\sigma}_{\Omega_i,n}^{2}>0$, this inequality is equivalently represented by the Schur-complement LMI
\begin{equation}
\begin{bmatrix}
c_{\Omega_i}
+
\bar{\sigma}_{\Omega_i,n}^{2}\gamma_{(l,l'),n}
&
\sqrt{c_{\Omega_i}}
\\
\sqrt{c_{\Omega_i}}
&
1-
\bar{\sigma}_{\Omega_i,n}^{2}
\xi_{\Omega_i,(l,l'),n}
\end{bmatrix}
\succeq\mathbf{0}.
\label{eq:app_scalar_lmi}
\end{equation}
When $\bar{\sigma}_{\Omega_i,n}^{2}=0$, the corresponding constraint reduces to $c_{\Omega_i}\xi_{\Omega_i,(l,l'),n}\leq\gamma_{(l,l'),n}$.
Define
\begin{equation}
\mathbf{S}_n
\triangleq
\mathbf{\Phi}_{n|n-1}^{-1}
+
\sum_{l=1}^{L}
\sum_{l'=1}^{L'}
\mathbf{J}_{(l,l'),n}^{T}
\operatorname{diag}
\left(
\boldsymbol{\xi}_{(l,l'),n}
\right)
\mathbf{J}_{(l,l'),n},
\label{eq:app_information_lower_bound}
\end{equation}
where
$\boldsymbol{\xi}_{(l,l'),n}=
[\xi_{\Omega_1,(l,l'),n},\ldots,
\xi_{\Omega_4,(l,l'),n}]^{T}$.
Because
$\operatorname{diag}(\boldsymbol{\xi}_{(l,l'),n})
\preceq
(\mathbf{V}_{(l,l'),n}^{\text{eff}})^{-1}$,
the information-filter expression in \eqref{eq:info_filter} gives
$\mathbf{S}_n\preceq\mathbf{\Phi}_n^{-1}$.
Since matrix inversion reverses the positive-semidefinite order, it follows that $\mathbf{\Phi}_n\preceq\mathbf{S}_n^{-1}$.

Introducing $\mathbf{A}_n$ as an upper bound on $\mathbf{S}_n^{-1}$, the Schur complement yields
\begin{equation}
\begin{bmatrix}
\mathbf{S}_n & \mathbf{I}_4\\
\mathbf{I}_4 & \mathbf{A}_n
\end{bmatrix}
\succeq\mathbf{0},
\label{eq:app_main_lmi}
\end{equation}
which implies $\mathbf{\Phi}_n\preceq\mathbf{A}_n$. Therefore, the original normalized tracking-error constraint is guaranteed by
\begin{equation}
\operatorname{tr}
\left(
\mathbf{T}^{-1}
\mathbf{A}_n
\mathbf{T}^{-T}
\right)
\leq
\operatorname{tr}(\overline{\mathbf{\Phi}})^{\max}.
\label{eq:app_trace_constraint}
\end{equation}

The reformulation is exact with respect to $\mathrm{C4}$. For any $\mathbf{W}_n$ satisfying the original constraint, selecting
$\xi_{\Omega_i,(l,l'),n}=
q_{\Omega_i,(l,l'),n}$
and $\mathbf{A}_n=\mathbf{\Phi}_n$ satisfies \eqref{eq:app_scalar_lmi}--\eqref{eq:app_trace_constraint}. Conversely, any point satisfying these LMIs yields $\mathbf{\Phi}_n\preceq\mathbf{A}_n$ and consequently satisfies the original tracking-error constraint. Hence, the reformulated and original constraints define the same feasible set in $\mathbf{W}_n$.

\section{Proof of Proposition~1}
Let $\mathbf{W}_n^\star \succeq \mathbf{0}$ be an arbitrary nonzero optimal solution of \text{P3}. For the considered $N_t$-element ULA, define the spatial-frequency steering vector as $\mathbf{a}(\omega)= \left[ 1,e^{j\omega},\ldots,e^{j(N_t-1)\omega} \right]^T$,
where $\omega=\pi\sin(\theta)$ for half-wavelength antenna spacing. The beampattern generated by $\mathbf{W}_n^\star$ is
\begin{equation}
G^\star(\omega)
=
\mathbf{a}^H(\omega)\mathbf{W}_n^\star\mathbf{a}(\omega)
=
\sum_{k=-(N_t-1)}^{N_t-1} c_k e^{jk\omega},
\end{equation}
where $G^\star(\omega)\geq0$ for all $\omega$ and $c_{-k}=c_k^*$. Thus, $G^\star(\omega)$ is a nonnegative trigonometric polynomial of degree at most $N_t-1$.
By the Fejér--Riesz theorem, there exists a polynomial $q(z)=\sum_{m=0}^{N_t-1} q_m z^m$ such that $G^\star(\omega)=|q(e^{j\omega})|^2$.

Define $\bar{\mathbf{w}}_n= [q_0^*,q_1^*,\ldots,q_{N_t-1}^*]^T$ and
$\bar{\mathbf{W}}_n= \bar{\mathbf{w}}_n\bar{\mathbf{w}}_n^H$.
Then, for every $\omega$,
\begin{equation}
\begin{aligned}
&\mathbf{a}^H(\omega)\bar{\mathbf{W}}_n\mathbf{a}(\omega)=
\left|\mathbf{a}^H(\omega)\bar{\mathbf{w}}_n\right|^2\\
& =
\left|\sum_{m=0}^{N_t-1}q_m e^{jm\omega}\right|^2
=
\mathbf{a}^H(\omega)\mathbf{W}_n^\star\mathbf{a}(\omega).
\end{aligned}
\label{eq:app_beampattern_preservation}
\end{equation}
Hence, the directional transmit power is preserved at every spatial frequency and, in particular, at all path departure angles.
The zeroth Fourier coefficient of $G^\star(\omega)$ satisfies $c_0=\operatorname{tr}(\mathbf{W}_n^\star)$.
Moreover, Parseval's identity gives
\begin{equation}
\operatorname{tr}(\bar{\mathbf{W}}_n)
=
\|\bar{\mathbf{w}}_n\|^2
=
\sum_{m=0}^{N_t-1}|q_m|^2
=
c_0
=
\operatorname{tr}(\mathbf{W}_n^\star).
\label{eq:app_trace_preservation}
\end{equation}
Since $\mathbf{W}_n^\star\succeq\mathbf{0}$ and $\mathbf{W}_n^\star\neq\mathbf{0}$, its trace is positive. Therefore,
$\bar{\mathbf{w}}_n\neq\mathbf{0}$ and
$\operatorname{rank}(\bar{\mathbf{W}}_n)=1$.
By assumption, the objective and constraints of \text{P3} depend on $\mathbf{W}_n$ only through its trace and the directional terms $\mathbf{a}_T^H(\theta_l)\mathbf{W}_n\mathbf{a}_T(\theta_l)$.
Consequently,
\eqref{eq:app_beampattern_preservation} and
\eqref{eq:app_trace_preservation} imply
\begin{align}
\hat{\Gamma}_n^C(\bar{\mathbf{W}}_n)
&=
\hat{\Gamma}_n^C(\mathbf{W}_n^\star),
\nonumber\\
\hat{P}_n^h(\bar{\mathbf{W}}_n,\rho_n)
&=
\hat{P}_n^h(\mathbf{W}_n^\star,\rho_n),
\nonumber\\
P_n^{\text{tot}}(\bar{\mathbf{W}}_n)
&=
P_n^{\mathrm{tot}}(\mathbf{W}_n^\star).
\label{eq:app_performance_preservation}
\end{align}

Furthermore, from the path-wise sensing SNR in \eqref{SensingSNR},
$\Gamma_{(l,l'),n}^{S}(\mathbf{W}_n)$ is proportional to $\mathbf{a}_T^H(\theta_l)\mathbf{W}_n\mathbf{a}_T(\theta_l)$.
Thus,
\begin{equation}
\Gamma_{(l,l'),n}^{S}(\bar{\mathbf{W}}_n)
=
\Gamma_{(l,l'),n}^{S}(\mathbf{W}_n^\star),
\qquad
\forall (l,l').
\end{equation}

The auxiliary variables
$\xi_{\Omega_i,(l,l'),n}^\star$
and
$\mathbf{A}_n^\star$
associated with
$\mathbf{W}_n^\star$
can therefore be retained. Since all path-wise sensing SNRs remain unchanged, constraints
$\widetilde{C}_4$--$\widetilde{C}_6$
remain satisfied. The transmit-power constraint is preserved by
\eqref{eq:app_trace_preservation}, while the communication and harvesting constraints are preserved by
\eqref{eq:app_performance_preservation}. Hence,
$\bar{\mathbf{W}}_n$
is feasible for \text{P3} and achieves the same objective value as
$\mathbf{W}_n^\star$.

Therefore,
$\bar{\mathbf{W}}_n$
is also optimal and satisfies
$\operatorname{rank}(\bar{\mathbf{W}}_n)=1$.
Hence, $\mathrm{P3}$ always admits an optimal rank-one solution, proving that the SDR is tight.
\hfill$\blacksquare$

\begin{remark}
Proposition~1 establishes the existence of an optimal rank-one solution, but an SDP solver may return a higher-rank optimum when the solution is nonunique. In that case, an equivalent rank-one solution can be recovered through Fejér--Riesz spectral factorization without loss of feasibility or optimality. Therefore, Gaussian randomization is unnecessary under the assumptions of Proposition~1.
\end{remark}

\begin{remark}
The tightness result relies on the ULA structure and on the dependence of $\mathrm{P3}$ on $\mathbf{W}_n$ only through its trace and directional beampattern values. It may not hold when additional constraints, such as per-antenna power limits or general transmit-covariance shaping constraints, are introduced.
\end{remark}

\bibliographystyle{IEEEtran}
\bibliography{References}
\vfill
\end{document}